\documentclass[aps,prd,twocolumn,preprintnumbers,superscriptaddress,floatfix]{revtex4}

\usepackage{multirow, graphicx,amssymb,url,mathrsfs,amsmath}
\usepackage{eucal,wrapfig,boxedminipage,setspace,subfigure}
\usepackage{amsxtra,amstext,latexsym,dsfont}
\usepackage[colorlinks]{hyperref}

\def\IR{{\hbox{{\rm I}\kern-.2em\hbox{\rm R}}}}
\def\IB{{\hbox{{\rm I}\kern-.2em\hbox{\rm B}}}}
\def\IN{{\hbox{{\rm I}\kern-.2em\hbox{\rm N}}}}
\def\IC{\,\,{\hbox{{\rm I}\kern-.59em\hbox{\bf C}}}}
\def\IZ{{\hbox{{\rm Z}\kern-.4em\hbox{\rm Z}}}}
\def\IP{{\hbox{{\rm I}\kern-.2em\hbox{\rm P}}}}
\def\IH{{\hbox{{\rm I}\kern-.4em\hbox{\rm H}}}}
\def\ID{{\hbox{{\rm I}\kern-.2em\hbox{\rm D}}}}

\def\det{{\rm det}}

\newcommand{\beq}{\begin{equation}}
\newcommand{\eeq}{\end{equation}}
\newcommand{\bea}{\begin{eqnarray}}
\newcommand{\eea}{\end{eqnarray}}

\begin{document}

\voffset 1cm

\newcommand\sect[1]{\emph{#1}---}

\title{Low $\mu$ and imaginary $\mu$ signals of a critical point in the phase diagram of an exactly soluble chiral symmetry breaking theory}

\author{Nick Evans}
\affiliation{ STAG Research Centre \&  Physics and Astronomy, University of
Southampton, Southampton, SO17 1BJ, UK}

\author{M.J. Russell}
\affiliation{ STAG Research Centre \&  Physics and Astronomy, University of
Southampton, Southampton, SO17 1BJ, UK}

\begin{abstract}

\noindent Holography has allowed the exact solution of a small number of large $N_c$ gauge theories. Amongst these is an ${\cal N}$=2 SYM theory of quarks interacting with ${\cal N}$=4 gauge fields. The temperature chemical potential phase diagram for this theory in the presence of a magnetic field is exactly known and shows first and second order chiral symmetry restoration transitions and a critical point. Here we extend this phase diagram to imaginary chemical potential to seek structure at small real $\mu$ and imaginary $\mu$ that help to reconstruct the large real $\mu$ phase structure.   We also explore a phenomenologically deformed version of the theory where the critical point can be moved into the imaginary chemical potential plane. In particular we observe that when the transition is second order  in these theories  there are naturally two distinct transitions - one for the onset of density and one for chiral symmetry restoration. In addition, the phase diagram has boundaries of regions where metastable vacua exist and these boundaries, as well as the phase boundaries, converge at the critical point. These observations may point to techniques for the study of the QCD critical point either on the lattice or using heavy ion collision data.
\end{abstract}

\maketitle

\newpage

\section{Introduction}

QCD displays a cross over phase transition  at finite temperature  (which we will represent by a second order transition in our massless theory) but it is widely assumed that the transition with chemical potential is first order. A critical end point should link the change in transition order \cite{Stephanov:1998dy}. The position of that critical point is a matter of considerable speculation but difficult to identify since the physics is non-perturbative and lattice Monte Carlo techniques are ineffective at large chemical potential (\cite{lattice} is a review of attempts to move to finite $\mu$ on the lattice).

Using the AdS/CFT Correspondence \cite{Maldacena:1997re,witten} a number of exact solutions of supersymmetric gauge theories have been found. Solutions also exist for less symmetric, deformed versions of those theories. Amongst these is an ${\cal N}$=2 SYM theory of a small number of quarks interacting with ${\cal N}$=4 gauge fields \cite{Karch:2002sh}. For the particular case of introducing a baryon number magnetic field \cite{Filev:2007gb}, which breaks the supersymmetry and conformal symmetry, the phase diagram is precisely known \cite{Evans:2010iy} (see the right hand side, real chemical potential, $\mu_R$, part of Fig 1). At low temperature, $T$, and density the preferred phase is characterized by a chiral symmetry breaking quark condensate  (formally a breaking of U(1)$_A$  since yukawa terms with the adjoint scalars break the SU($N_f)_A$ to  U(1)$_A$ - that U(1)$_A$ is a good symmetry at large $N_c$).

The model does not have confinement. However, it is possible in QCD that confinement is a property of the pure glue theory below the IR mass of the quarks induced by chiral symmetry breaking. Then the chiral transition is the key dynamics as in the solved model. 

In the holographic model we study the temperature transition is first order whilst the transition with chemical 

\begin{center}
\includegraphics[scale=0.4]{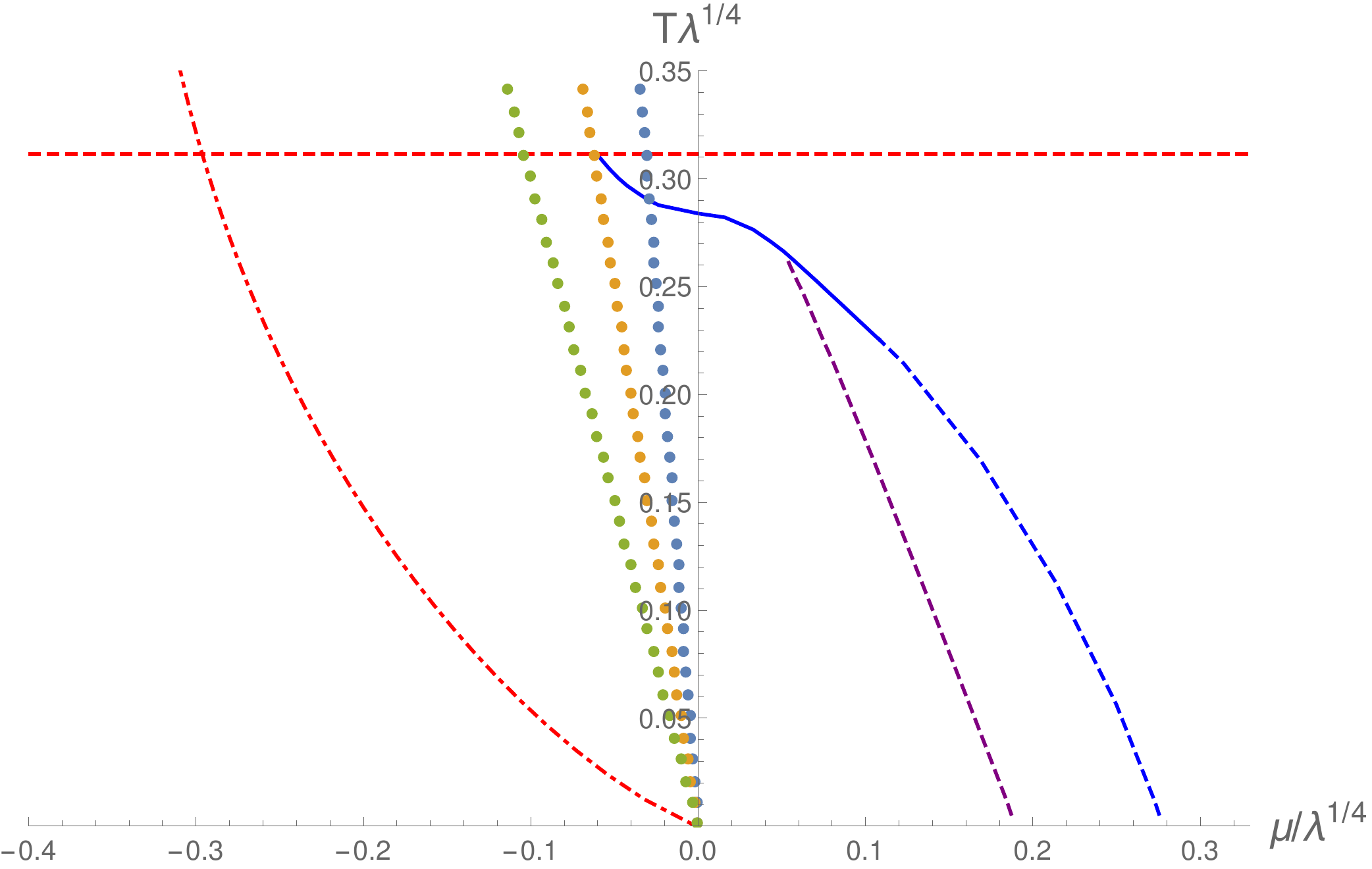} 

\noindent {\it Figure I:  The phase diagram of the ${\cal N}=2$ theory with a magnetic field. T and $\mu$ here are expressed in units of the magnetic field $\sqrt{B}$. The positive $\mu$ axis is real $\mu_R$ whilst negative $\mu$ correspond to imaginary $\mu_I$ values.  The blue line is the chiral restoration transition (solid is first order, dashed second order); the purple line is the second order transition associated with the onset of density. The horizontal red dashed line shows where the chirally broken vacuum ceases to be a turning point of the effective potential - in the imaginary $\mu_I$ plane above this line the effective potential is unbounded. The curved red dashed line shows where vacua with density becomes unstable - to the left in the imaginary $\mu_I$ plane there are again instabilities. The dotted lines show the positions of the Roberge Weiss transition for  $N_c \lambda^{1/2}=10\pi$ (blue), $5\pi$ (orange) and $3\pi$ (green).}
\end{center} \vspace{-0.4cm}

potential is continuous (in fact it splits into two  
 continuous transitions one at which the mesons of the theory  melt \cite{Hoyos:2006gb} and density switches on and a second at which chiral symmetry is restored). There are critical points for each transition. The theory, though distinct in detail from QCD,  at least 
has some of the generic features of interest.

In this paper we want to take this theory  and ask how clearly, if at all, can we identify the position of the critical point from the study of the phase diagram at imaginary chemical potential, $\mu_I$, low values of real chemical   potential, $\mu_R$, or from isolated data points as if from heavy ion collision data. The hope is that by asking these questions in a solved theory we might generate new ideas that might apply to QCD. 

Our first job is to extend the phase diagram of the theory to the  $\mu_I - T $ plane. We will focus on the massless quark theory so that second order lines are precise. 

A previous analysis \cite{Aarts:2010ky} of the D3/D7 system (without a magnetic field) has concentrated on the Roberge-Weiss transitions \cite{Roberge:1986mm} of such theories (see also the holographic work in \cite{Rafferty:2011hd,Bigazzi:2014qsa,Isono:2015uda} and most recently \cite{Ghoroku:2020fkv}). Here the key physics is that a spurious U(1)$_B$ transformation with parameter $\alpha=\mu_I x$ can remove the chemical potential from the action. The quark fields are rotated by $e^{i \alpha}$ though so baryonic operators have a discontinuity in their boundary conditions around the thermal circle. In the case where the resulting phase difference is a multiple of $2 \pi T/N_c$ (with $N_c$ the
number of colours) a gauge transformation that differs around the thermal circle by an element of the centre of the group can be used to remove $\mu_I $ completely. The $\mu_I=0$ and $\mu_I=2 \pi T/N_c$ theories are therefore identical.  The result is that there must be first order transitions at 
\begin{equation}\mu_I/T = (2k+1)\pi/N_c, \hspace{1cm} k=0,1,2...\end{equation} 
At very large $N_c$ these become very dense and begin essentially at $\mu_I=0$.  Our hope here though is that at lower $N_c$ near, for example, $N_c=3$ they become less dense and pushed out to large $\mu_I$ so they can be neglected. Nevertheless we hope that $N_c=3$ is close enough to large $N_c$ that aspects of our analysis remain useful. In particular in Figure I the first transition occurs on the line 
\begin{equation} \lambda^{1/4} T= {N_c \lambda^{1/2} \over \pi } ~~ {\mu\over \lambda^{1/4}} \end{equation}
We have plotted the transition line for $N_c \lambda^{1/2} = 10 \pi, 5 \pi$ and $3 \pi$ in Fig 1 and for $N_c \lambda^{1/2} \leq 5\pi$ all the physics we will use is present. This is still  strong coupling.

In practice we just concentrate on the role of $\mu_I$ in the DBI action for the probe branes in Schwarzschild AdS$_5$ describing the quarks. The result is shown on the left in Fig I - the first order transition extends a little way into the $\mu_I$ piece of the $\mu-T$ plane before the theory becomes unstable. We have checked that the transition line is linear in $\mu^2$ across the $\mu=0$ axis as one would expect. Instabilities exist at $\mu_I$ to the left or above the red dotted lines in Fig I as we will discuss.  Again if $N_c \lambda^{1/2}>5\pi$ then the Roberge Weiss transition occurs before any of the instabilities set in and this may indicate that this is the smallest value of $N_c \lambda^{1/2}$ compatible with

\begin{center}
 \includegraphics[scale=0.4]{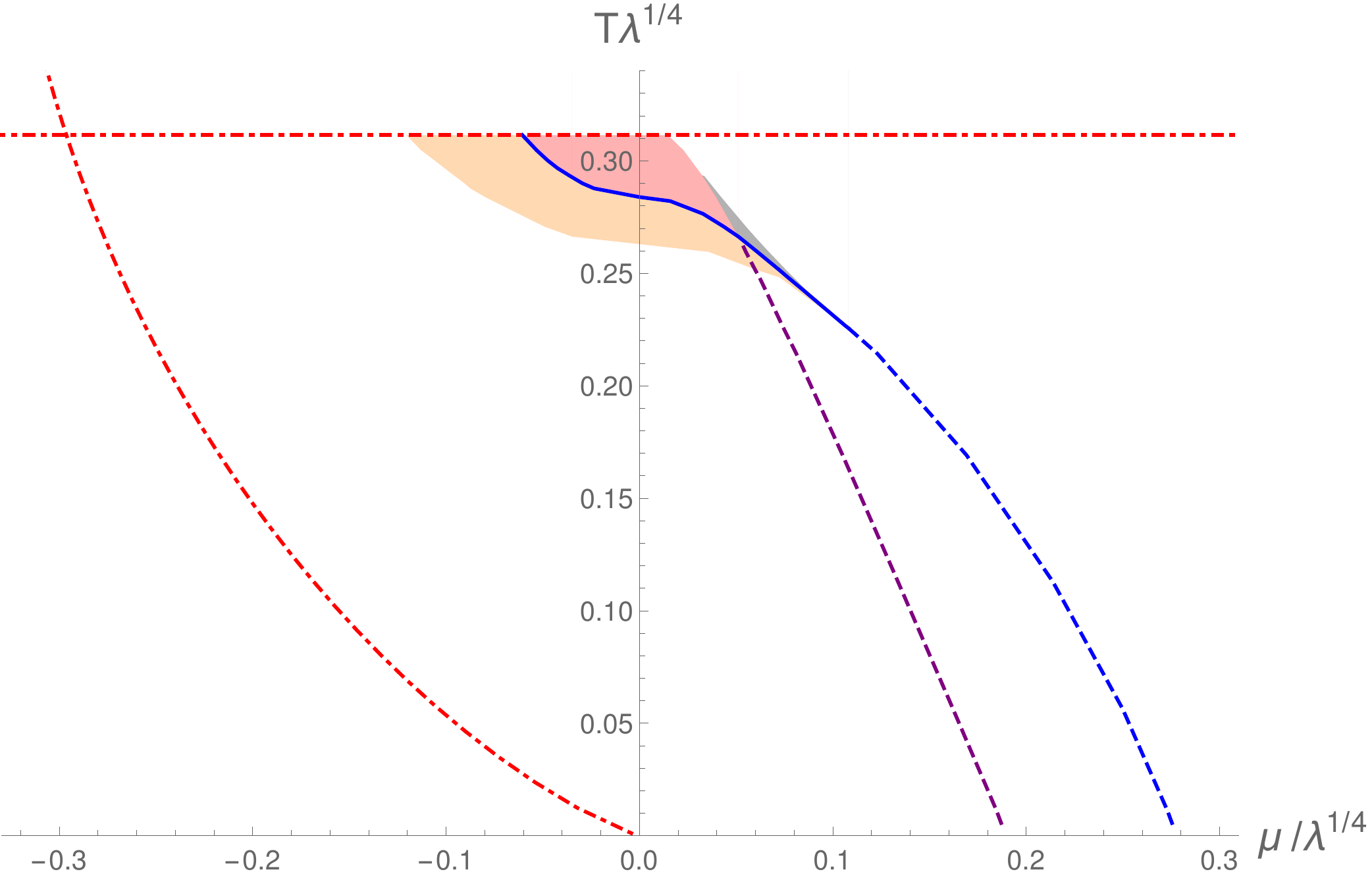}

\noindent {\it Figure II:  The phase diagram of the   $ {\cal N}=2$  theory with B field from Fig I but now in addition showing regions with metastable vacua. In the orange shaded region the chirally symmetric vacuum is metastable. In the red region the chirally broken vacuum with zero density is metastable. There is also a small grey region where a chirally broken, dense state is metastable which is addressed carefully in Section III. The boundaries of the metastable regions and the transition lines themselves converge close to the region with the critcal points - the shaded region ``points'' to the critical point region. }
\end{center} \vspace{-0.4cm}

the large $N_c$ analysis - this limit is sufficient for our purposes. As shown, at this stage, there is little to be deduced for the real chemcial potential, $\mu_R$, region.

The theory, however, contains more information than just the ground state. We can find all turning points of the effective potential and it's interesting to track these and show in which regions of the parameter space there are metastable vacua (also sometimes called spinoidal regions - for example in \cite{Nickel:2009wj,Kovensky:2019bih}). The metastable states exist in a band around the first order transition. This is shown in Fig II - for the moment the reader should just associate the shaded regions with the presence of metastable vacua. On the outer edges of this band the peak in the effective potential between the true vacuum and the false vacuum merge with the false vacuum (there 
is some structure to this boundary in this model that we will elaborate on below) as the metastable vacuum disappears. These boundaries, which are distinct from the first order transition line, smoothly become the second order transition line(s) at the critical points. There are therefore actually three lines we can draw which must converge at, or near, the critical point. If one can identify these lines at finite $\mu_I$ or at low $\mu_R$, as one can here, then extrapolation provides a sensible guess to the position of the critical point (one could reasonably estimate the $T,  \mu$ values at the critical point at the 10$\%$ level).

Having just a single theory, of course, makes it hard to learn generic lessons but equally fully solvable models are scarce. Previously \cite{Evans:2011eu} we have applied some ``bottom 
up" parameters to this model that allows us to move the position of the critical point and even change the order of the phase transitions. An example, that we will use here, is a parameter in the black hole emblackening factor  which distorts the horizon from a sphere to an  ellipse. Whilst this is not a full solution of the supergravity  equations it does at least encode the breaking of the symmetry between the directions parallel and perpendicular to the D7 branes so may be indicative of the behaviour of a backreacted D3/D7 solution. The parameter can be used to move the chiral symmetry breaking critical point towards $\mu_R=0$ and we show here it can even push it  through the axis into the $\mu_I$ plane leaving the pure temperature transition  second order.  We repeat our study in some of these cases to show how generic our conclusions are and to display some other possible structures. 

In our final section we will try to draw lessons on possible structures in both generic phase diagrams  and for QCD.  We speculate as to whether regions of the QCD phase diagram accessible to computation on the lattice might contain metastable vacua, in which case the boundarries of these regions could be used to point to the critical point. Also the second order transition line may separate in the $\mu_I$ plane into several transitions, including one for the onset of density and another for chiral symmetry restoration - these two lines might point to the critical point. In practice these transitions will be blurred into the crossover transition though and are likely very hard to spot even if they exist.

Heavy ion collision data might also be able to identify regions of the phase diagram with metastable vacua. In such regions there might be events in which the vacuum becomes caught for a period in the metastable state. It is possible that such states will hadronize differently and form an identifiably distinct set of events indicating that the theory is in a region with metastable vacua. This again might help distinguish boundaries of the regions with  metastable vacua. See \cite {Gazdzicki:2020jte} for a recent summary of heavy ion collision searches for the QCD critical point. \vspace{-1cm}

\section{The Holographic Description}  \vspace{-0.5cm}

Using D7 brane probes \cite{Karch:2002sh} to introduce quarks to the basic D3 brane AdS/CFT Correspondence is now a well established technique. At zero temperature we use an AdS$_5 \times$ S$^5$  geometry with coordinates
\begin{equation} ds^2 = 
{r^2 \over R^2} dx_{4}^2 + {R^2 \over r^2} \left( d\rho^2 +
\rho^2 d\Omega_3^2 + dL^2 + L^2 d\Omega_1^2\right)
\end{equation}
where we
have split the coordinates into the $x_{3+1}$ of the gauge theory,
the $\rho$ and $\Omega_3$ which will be on the D7 brane
world-volume and two directions transverse to the D7, $L, \phi$.
The radial coordinate, $r^2 = \rho^2 + L^2$, corresponds
to the energy scale of the gauge theory and radius of the space is  $R^4 = 4 \pi g_{uv}^2 N_c \alpha^{'2}$.

We will introduce a D7 probe brane  into the
geometry to include quarks - the probe approximation is equivalent to
working in a quenched approximation. This system has a U(1)$_A$ axial symmetry on
the quarks, corresponding to rotations in the angle $\phi$,
which will be broken by the formation of a quark condensate.

We seek D7 embedding functions  $L(\rho)$ at some fixed $\phi$. The Dirac Born
Infeld action is 
\begin{equation}
\begin{array}{ccl}
S_{D7} & = & - N_f T_7 \int d^8\xi e^\phi  \sqrt{- \det (P[G]_{ab}+ 2 \pi \alpha' F_{ab})}\\ &&\\
&=&  - N_f \overline{T_7} \int d^4x~ d \rho ~ \rho^3 \beta \sqrt{1 +
(\partial_\rho L)^2} \end{array} \end{equation}
 where $T_7 = 1/(2 \pi)^7
\alpha^{'4}$ and $\overline{T_7} = 2 \pi^2 T_7/g_s$ after integrating over the 3-sphere on the D7. The factor of $\beta$ appears when a magnetic field is introduced through eg $F_{12}=B/2 \pi \alpha'$ \cite{Filev:2007gb} \vspace{-0.9cm}

\begin{equation} \beta = \sqrt{1 + {B^2 R^4 \over (\rho^2 + L^2)^2}} 
\label{Bform} \end{equation}
Note that it enter as an effective dilaton term although it's origin is in the DBI action.

The equation of
motion for the embedding function is therefore 
\begin{equation} \label{embed}
\partial_\rho \left[ {\beta \rho^3
\partial_\rho L \over \sqrt{1+ (\partial_\rho L)^2}}\right] - 2 L \rho^3
\sqrt{1+ (\partial_\rho L)^2} {\partial \beta \over \partial
r^2} = 0 \end{equation}
The UV asymptotic of this equation has solutions of the form   \vspace{-0.9cm}

\begin{equation}
\label{asy}L = m + {c \over \rho^2} +... 
\end{equation}
where we 
interpret $m$ as the quark mass ($m_q = m/2 \pi \alpha'$) and $c$
is proportional to the quark condensate.

There is always a solution $L=0$ which corresponds to a massless
quark with zero quark condensate ($c=0$). However, for forms of $\beta$ such as that in (\ref{Bform}) which grow near the origin there are symmetry breaking solutions that have $m=0$ in the UV but bend off axis (at a particular, symmetry breaking, value of $\phi$) to end on the $L$ axis with $L'(0)=0$. These ``Minkowski`` embeddings are the minimum of the effective potential (computed by evaluating minus the action on the solution). The $L=0$ embedding is a local maximum of the potential.

Temperature can be included in the theory by using the AdS-Schwarzschild 
black hole metric as proposed by Witten \cite{witten}. 
The metric is
\begin{equation} 
ds^2 =  -{K(r) \over R^2} dt^2 + {R^2 \over K(r)}dr^2+ \frac{r^2}{R^2} d\vec{x}_3^2
+ R^2 d\Omega_5^2
\end{equation}
\begin{eqnarray}
 K(r) = r^2-\frac{r_H^4}{r^2} \ , \qquad r_H := \pi R^2 T \ .
\end{eqnarray}
$r_H$ is a dimension one parameter identified with temperature T.

It is helpful to make the coordinate transformation \cite{Babington:2003vm}
\begin{eqnarray}
  \frac{rdr}{(r^4-r_H^4)^{1/2}} \equiv \frac{dw}{w}
   \\ \ \ \  \  2w^2 = r^2 + \sqrt{r^4 - r_H^4}\ , 
\end{eqnarray}
with $\sqrt{2}w_H = r_H$. The metric becomes
\begin{equation} \begin{array}{ccl}
   ds^2 & =&   \frac{w^2}{R^2}(- g_t dt^2 + g_x d\vec{x}^2)  \\
   &&\\ &&  + \frac{R^2}{w^2} (d\rho^2 + \rho^2 d\Omega_3^2
         + dL^2 + L^2 d\Omega_1^2 ), \end{array} 
\end{equation}
where
\begin{eqnarray}  \label{schwarschild}
g_t = \frac{(w^4 - w_H^4)^2}{ w^4 (w^4+w_H^4)}\ ,  \qquad
g_x  = \frac{w^4 + w_H^4}{ w^4} \ .
\end{eqnarray}
\begin{eqnarray}
  w = \sqrt{\rho^2 + L^2}\ ,  \quad \rho = w \sin\theta \ .
  \quad L = w \cos\theta \ ,
\end{eqnarray}
The Lagrangian for the magnetic field case becomes
\begin{equation} \begin{array}{ccl} {\cal L} & = & -\overline{T_7} \rho^3 \left( 1 - {w_H^4 \over w^4} \right) \sqrt{1 + (\partial_\rho L)^2} \\
&& \left. \right. \hspace{1cm} \times \sqrt{\left( 1 + {w_H^4 \over w^4} \right)^2 +{R^4 B^2 \over w^4} } \end{array}\end{equation}
The embedding equation for $L(\rho)$ is straightforward to derive. Minkowski embeddings exist until the black hole horizon ``eats" the central area of the $\rho-L$ plane. The flat $L=0$ embedding always exists and so there is a first order transition from Minkowski to flat at a critical value of T \cite{Babington:2003vm,Mateos:2006nu}.

A chemical potential is introduced through the U(1) baryon number gauge field $A_t$ component \cite{Kobayashi:2006sb}
which enters the DBI action as  
\begin{equation} \label{muL} \begin{array}{ccc}
{\cal L} & = &  -\overline{T_7}\rho^3 \left( 1 - {w_H^4 \over w^4} \right) \sqrt{\left( 1 + {w_H^4 \over w^4} \right)^2 +{R^4 B^2 \over w^4} }\\
&&\\
&& \sqrt{1 + (\partial_\rho L)^2) -  {w^4( w^4 + w_H^4) \over (w^4 - w_H^4)^2} (2 \pi \alpha' A_t)^2   } \end{array}  \end{equation}
There is a conserved quantity $d$ (the density) associated with $A_t$. We can Legendre transform the action to write $A_t$ in terms of $d$ leaving (after rescaling all dimensionful objects to be in units of $R \sqrt{B}$ donoted by the tildes) 
\begin{equation} \label{tmuL}\tilde{\cal L} = -\overline{T_7} {\tilde{w}^4 -\tilde{w}_H^4 \over \tilde{w}^4} \sqrt{K (1 + (\partial_{\tilde{\rho}} \tilde{L})^2) }\end{equation}
\begin{equation} K = \tilde{\rho}^6 \left( {\tilde{w}^4 +\tilde{w}_H^4 \over \tilde{w}^4} \right)^2 + {\tilde{\rho}^6 \over \tilde{w}^4} + {\tilde{w}^4 \tilde{d}^2\over \tilde{w}^4 + \tilde{w}_H^4} \label{kay}\end{equation}
Given a solution for L at some $T,d$ one can then find the chemical potential as
\begin{equation} \label{mud}
\tilde{\mu} = \tilde{d} \int_{\tilde{\rho}_H}^\infty d \tilde{\rho}  {\tilde{w}^4 - \tilde{w}_H^4 \over\tilde{w}^4 + \tilde{w}_H^4} \sqrt{{1 + (\partial_{\tilde{\rho}} \tilde{L})^2 \over K}} \end{equation}

At small T the appropriate solutions as $d$ begins to grow from zero are solutions that end on the black hole horizon at the origin but ``spike" up to the form of the Minkowski embedding. There is a corresponding non-zero critical $\mu$ for the on-set of $d$. There is a continuous transition here as the Minkowski embedding becomes a black hole ending embedding. As $d$ then increases the black hole solution smoothly evoles to merge with the flat $L=0$ embedding in a second continuous transition (where chiral symmetry breaking switches off) at a higher critical $\mu$.

The full $\mu - T$ phase diagram is discussed in detail in \cite{Evans:2010iy}. Here we use two techniques to find the transition lines that will interest us: 

1) To locate first order transitions: At a fixed $T,d$ we seek Minkowski embeddings and then evaluate the difference in free energy between these and the L=0 embedding. We then vary $d$ to locate the first order transition point where these embeddings are degenerate in energy. One then repeats at all T.

2) To locate second order transitions: at fixed $T,d$ we find embeddings shooting off the black hole surface from an angle $\theta$ and read off the UV asymptotic value of $m$. Now varying $d$ we seek points where massless solutions merge with the flat embedding at $\theta=\pi/2$ or the Minkowski embedding at $\theta = 0$ or points where two new solutions emerge. Again one repeats at all T.

Imaginary chemical potential solutions are found by simply allowing $A_t \rightarrow iA_t$ or $d \rightarrow id$ and repeating the process.  The lagrangian with  imaginary chemical potential is unchange except in the factor $K$ in equation (\ref{kay}) where  $d^2 \rightarrow - d^2$.

In Appendix A we present detailed computations and plots for one particular T slice across the phase diagram including evaluating the free energy of the solutions.

 It is worth pausing to write the phyiscal temperature and chemical potential in terms of $\mu$, $w_H$ and the physical B field that emerge from (\ref{tmuL}) and (\ref{mud}). We have
 \begin{equation} T_{\rm phys} = {\sqrt{2} \tilde{w}_H \over \pi R^2} R \sqrt{2 \pi \alpha' B_{\rm phys}} = { 2 \tilde{w}_H \over \sqrt{\pi} \lambda^{1/4}} \sqrt{B_{\rm phys}} \end{equation}
\begin{equation} \mu_{\rm phys} = {\tilde{\mu} \over 2 \pi \alpha'} R \sqrt{2 \pi \alpha' B_{\rm phys}} = { \tilde{\mu} \lambda^{1/4} \over \sqrt{2 \pi} } \sqrt{B_{\rm phys}} \end{equation}
Note that  these are independent of $\alpha'$ as they must be since $\alpha'\rightarrow 0$ in the supergravity limit. In our plots we plot $\mu_{\rm phys} / \lambda^{1/4}$ against $\lambda^{1/4} T_{\rm phys}$ and set $\sqrt{B_{\rm phys}}=1$.
\vspace{-0.7cm}

\section {The Phase Structure} \vspace{-0.5cm}

We show the phase transition structure for the model in Fig I (the positive $\mu$ axis is $\mu_R$ the negative axis $\mu_I$). In the phase including $T = \mu = 0$ the vacuum is characterized by chiral symmetry breaking and zero density (it is a so called Minkowski embedding in the brane picture). At high $T, \mu_R$ the vacuum is a chirally symmetric state with generically melted mesons and non-zero density (these are flat embeddings). At low $\mu_R$ there is a first order thermal transition between these vacua as T grows. At larger $\mu_R$ there is a region with a third low $T$ vacuum which has a density of deconfined quarks but which are still massive due to chiral symmetry breaking (a black hole embedding). The chiral restoration transition has a critical point where the transition becomes second order. The transition from the low $T,\mu_R$ phase to the deconfined massive quark phase is second order. The transition from that phase to the chirally symmetric phase is also second order.

Our first new results here are that we have extended the phase structure to imaginary chemical potential, $\mu_I$. This is shown on the left hand side of Fig I. Note we have checked in all our figures to come that the transition line  is linear in $\mu^2$ across the $\mu=0$ axis as one would expect. The first order chiral transition extends into the imaginary $\mu$ plane before coming to a halt on the line at T=0.311 (shown by a red dashed line in Fig I). Above this value of $T$ the remnant of the low $T, \mu$ chiral symmetry breaking vacua ceases to be a turning point of the effective potential. The Minkowski embedding can not exist above this value of $T$ because the black hole is too large and blocks the IR of the solution. Note this is independent of $\mu$ since the Minkowski embedding simply has $A_t=\mu$ which does not contribute to the action. At real $\mu_R$ this state is no longer the true vacuum  and this line simply marks where the state ceases to be a turning point of the potential. At larger $\mu_I$ though below the T=0.311 line the Minkowski embedding (chiral symmetry breaking state) is the true vacuum yet it suddenly vanishes at higher T. The only explanation is that at higher T, for these $\mu_I$, the effective potential becomes unbounded at large values of the condensate. Thus above the line $T=0.311$ the theory is ill-defined for $\mu_I$ - we will henceforth not consider temperature above that value.

In Fig 1 there is a second dashed red line emerging from the origin and cutting the $T-\mu_I$ plane. To the left of this line the flat or black hole embeddings do not exist because the action turns imaginary . 
To make this explicit consider the parameter $K$ of (\ref{kay}) which is square rooted in the action (\ref{muL}) - we can evaluate it near the black hole horizon. We set $w_H \sim T$ and remember $\omega^2 = \rho^2 + L^2 $, where at the black hole horizon we have $\rho \sim T \sin\theta$ and $L \sim T\cos\theta$, we have
\begin{equation}
\begin{array}{ccc}
 K &=& (T \sin\theta)^6 \left( {((T \sin\theta)^2 + (T\cos\theta)^2)^2 + T^4 \over ((T \sin\theta)^2 + (T\cos\theta)^2)^2} \right)^2 \\ &&\\
 &&+{(T \sin\theta)^6 \over ((T \sin\theta)^2 + (T\cos\theta)^2)^2} \\  &&\\
 && - {((T \sin\theta)^2 + (T\cos\theta)^2)^2 d^2\over ((T \sin\theta)^2 + (T\cos\theta)^2)^2 + T^4}
\end{array}
 \end{equation}
The final negative term is trying to force the action to become imaginary. The first two positive terms both go as $\sin\theta$ so  for a given temperature increasing density forces the minimum value of $\theta$ to take on larger values. For large enough $d$ this value becomes $\pi/2$ and above this value of $d$ there can be no further flat or black hole solutions. The curved red dotted line in Fig I is where this criteria is met. 
This corresponds to another instability of the effective potential against moving to larger density. Thus our analysis will be restricted to the part of the $\mu_I$ plane bounded by the red dashed lines in Fig I. We note that very close to this boundary there are some additional black hole solutions but we will not investigate these further since we wish to focus around the critical point.  As we mentioned  in the introduction if $N_c\lambda^{1/2}\geq5\pi$ then the instability regions are not part of the true vacuum of the theory because to the left of the Roberge Weiss transitions there are just repeats of the phyiscs to the right of the transition line in Fig I. This may indicate that this value of $N_c\lambda^{1/2}$ is the minimum possible value compatible with the large $N_c$ limit - that minimum value is sufficient for our discussions here.

Let us now imagine that some theorist can compute at imaginary $\mu_I$ and only very low  $\mu_R$ values and they are hoping to understand the large $\mu_R$ structure to see if there are one or more critical points in this case. Looking at Fig I there is little to guide this theorist - he could perform a fit to the first order chiral transition line as it crosses the T axis and perhaps do a reasonable job of predicting where the transition contour is at real $\mu$ but there is  apparently no hint as to where the critical point must lie.

To gain more insight our putative theorist could make use of more information that is available to him. In particular the first order transition is associated with a crossing of two distinct vacua and to either side one or the other is a metastable vacuum state. We can ask the question where are there metastable vacua in the plane?

The answer in this case is illustrated in Figure II. Here the orange region is where the chirally symmetric vacuum (flat embedding) is metastable. In the red shaded region the Minkowski embedding (chirally broken, $d=0$ phase) is metastable.  In the small gray region a black hole embedding (chirally broken but $d \neq 0$) is metastable.

\newpage

\begin{center}
\includegraphics[scale=0.55]{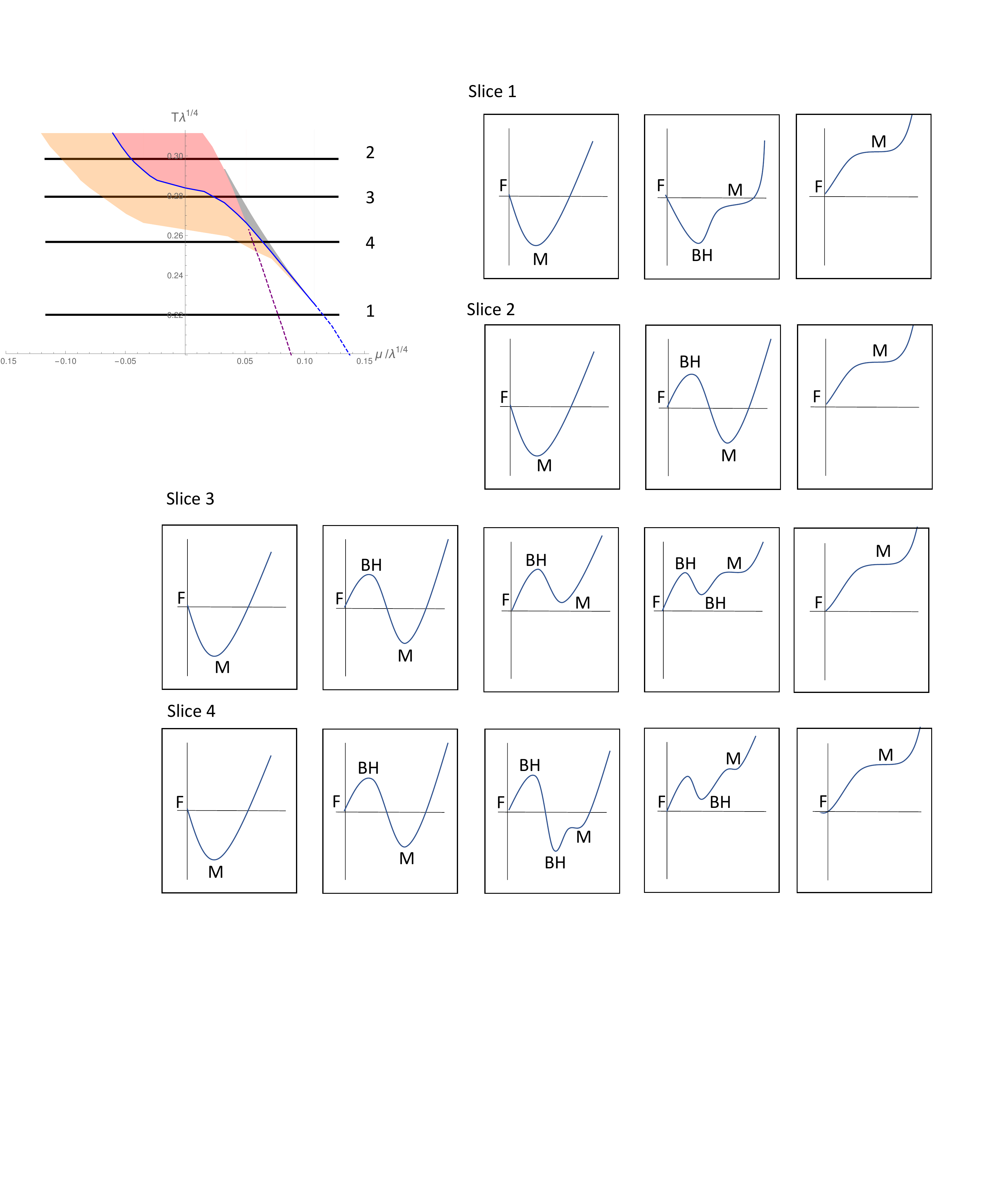} \vspace{-6cm}

\mbox{ \it Figure III: The structure of the effective potential against quark condensate as one moves from left to right across }
\mbox{ \it the phase diagram of the B field theory on a number of different T slices. Turning points of the potential are  }
\mbox{ \it  marked dependent
by their nature: F (flat, chirally symmetric); BH (black hole, dense massive phase); }
\mbox{\it M (Minkowski,  chirally broken)}
\label{slices}
\end{center}  

To understand this picture better it is helpful to use it to reconstruct the effective potential of the model across a number of fixed T slices - see Figure III where we zoom 
in on the interesting structure.

Slice one is at low temperature: starting at reasonably large imaginary $\mu_I$ the chiral symmetry breaking (zero density) phase is preferred (Minkowski embedding). As we track right a second order transition occurs to the deconfined massive quark phase (a black hole

\newpage
$\left. \right.$  \vspace{18.5cm}

embedding emerges from the Minkowski embedding) - here we believe the chiral symmetry breaking vacuum turns into a point of inflection of the effective potential.  Then sequentially a second order transition to the chirally symmetric (flat embedding) occurs - that is the black hole embedding merges with the flat embedding and then ceases to exist. Note here at no stage are there metastable vacua (we assume a point of inflection is insufficient to be  visible either through lattice studies or by noticeable events in a heavy ion collider).

Moving now to large $T$ - slice 2. To the left again the chirally broken vacuum dominates. As we track left to right the first event is that a peak  in the effective potential (black hole embedding) emerges from the chirally symmetric state as that state becomes a local minimum metastable vacuum. The chirally broken and chirally symmetric vacuum then interchange at a first order transition leaving the chirally broken phase as the metastable vacuum. The final change is that the chirally broken vacuum and the maximum of the potential merge to again leave the chirally broken phase remnant as a point of inflection.

Slices 3 and 4 show some more subtle structure around the critical points. Slice 3 follows slice 2 (moving to the 
right) until after the first order transition. Now the 
chirally broken remnant does not just merge with the potential peak but converts itself to a point of inflection throwing off a potential minimum that then moves to merge with the potential peak. In this intermediate region there is a metastable vacuum which is a deconfined massive quark phase (black hole embedding). Typically these minima are less deep than when the chirally broken or symmetric vacua are metastable. We show regions of the phase diagram with these metastable vacua in grey.

Slice 4 shows a further mixing of these events - the 
chirally broken vacuum converts to a deconfined massive quark phase at a second order transition before the first order transition. Thus the resulting first order transition is from the deconfined massive quark phase to the chirally symmetric phase. 

Hopefully this slice analysis has helped the reader to interpret the figures we present.

 A key observation is to follow the behaviour of the two boundaries where the chirally symmetric and chirally  broken  phase become metastable at high temperature - the boundaries of the red and orange regions. At each of these a black hole embedding merges with either the flat or Minkowski embedding. As one moves to lower T these boundaries become precisely the second order phase lines 
where again black hole embeddings merge with flat and Minkowski embeddings.  
Necessarily these boundaries of the metastable vacua region must join the transition lines at the critical points where the order of the transitions change. Further that these boundaries are distinct at high T naturally transforms to them being distinct at low T producing the three phases and  two second order transitions we see.

The interesting  thing about Fig II in comparison to Fig I is that whilst the latter simply has transition lines that apparently randomly convert their order, Fig II essentially has arrows pointing to the critical points! In particular the boundary where the chirally broken phase becomes metastable and the transition line itself converge 
at the critical point for chiral symmetry breaking. Equally the  boundary where the chirally symmetric 
phase  becomes metastable and the transition line converge is the critical point on the transition line to the deconfined quark phase. Thus our putative theorist who
can only compute at imaginary $\mu_I$ and low $\mu_R$ relative to T could try to identify the edges of the metastable regions and then extrapolate them to make an approximation as to the  positions of the critical points.

If the theorist could also access heavy ion collision data from a variety of experiments then he could hope to identify whether in those experiments $\mu, T$ lie in the 
metastable regions - for example one might expect two different categories of events, one which is ignorant of the metastable vacua and one of which got stuck in the metastable state for a period. These events could plausibly have different signatures even after hadronization. One might be able to further map out the metastable region even if the critical point had not been hit directly.   \vspace{-0.6cm}

\section{Horizon Deformed Theories}  \vspace{-0.4cm}

So far we have only considered a single phase structure so one might wonder how generic any of the features we see are. In principle it would be good to study many other such holographic set ups yet the number of fully understood ones are few and far between. In the future  it would be interesting to construct more eg based on the D4/D6 system \cite{Kruczenski:2003uq}. For now,  we will add a ``bottom-up''  parameter to our probe D3/D7 system with a B field which was first introduced in \cite{Evans:2011eu}. Of course ``bottom-up'' is synonymous for an incomplete model but the trick we use is instructive and allows us to rather simply move the critical point. 

The trick is to deform the spacetime geometry by making the substitution
\begin{equation}
w^2 \rightarrow \rho^2 + {1 \over \alpha} L^2, \hspace{1cm}\alpha > 1
\end{equation}
into the metric factors in (\ref{schwarschild}). This is not a solution of the Einstein equations but it breaks the $\rho-L$ symmetry which at least would happen were one to backreact the D7 branes (they are extended in $\rho$ but point like in $L$). More practically by squashing the black hole horizon onto the $L$ axis it is harder for the black hole to disturb the Minkowski embeddings that describe the chirally broken vacuum. This change tends to favour second order chiral transitions with temperature. The test  of the use of this approach is rather in what one learns by doing it. In Figure IV we display (zoomed to the interesting segments) phase diagrams for a variety of $\alpha$'s.

If one first concentrates on the phase lines then the effect of growing $\alpha$ is  to push the critical point on the chiral line towards the T axis. At $\alpha=1$ the critical point on the chiral transition line lay after the separation  of
the 

\newpage

\begin{center}

\mbox{\includegraphics[scale=0.34]{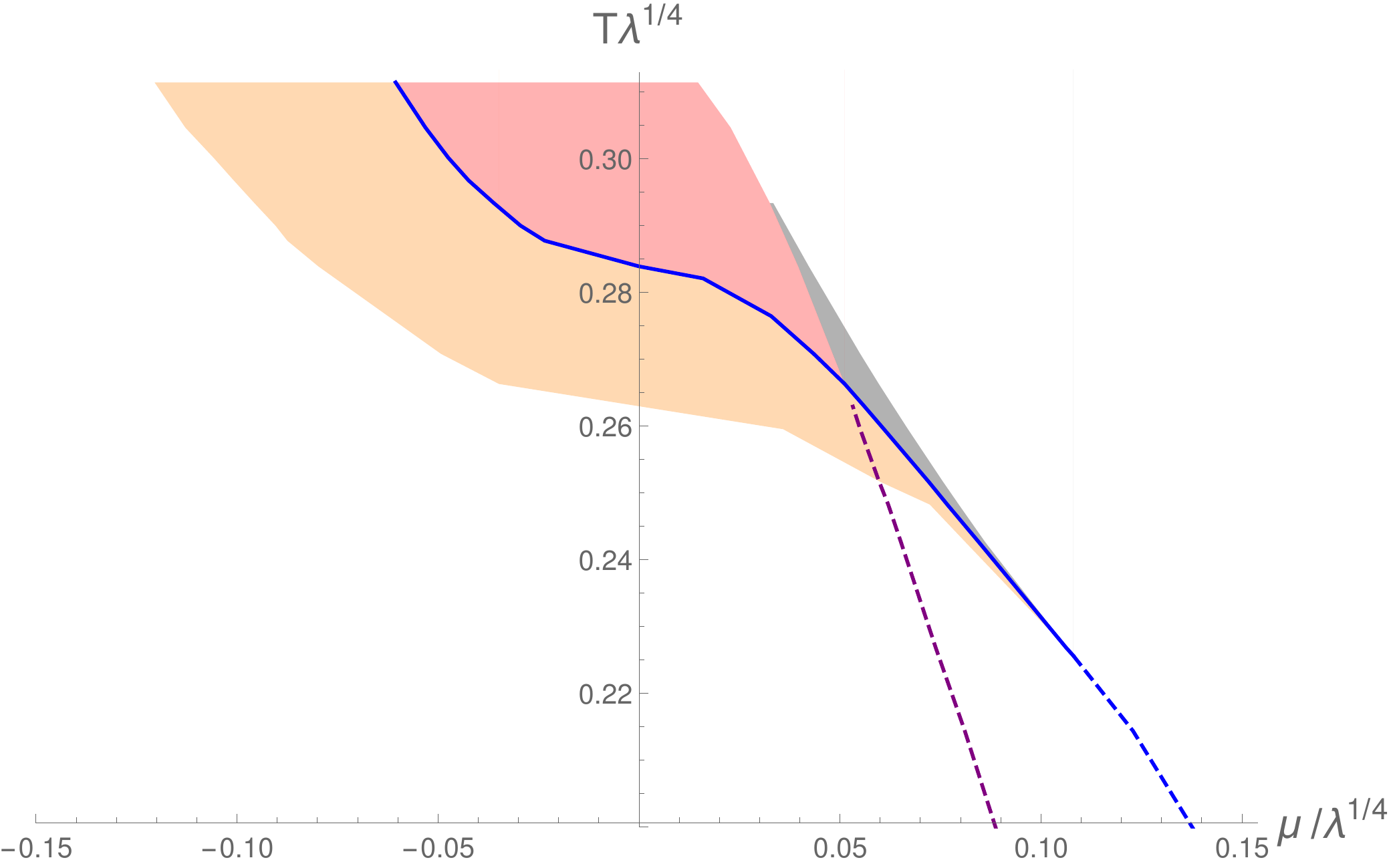}  \hspace{1.5cm} \includegraphics[scale=0.34]{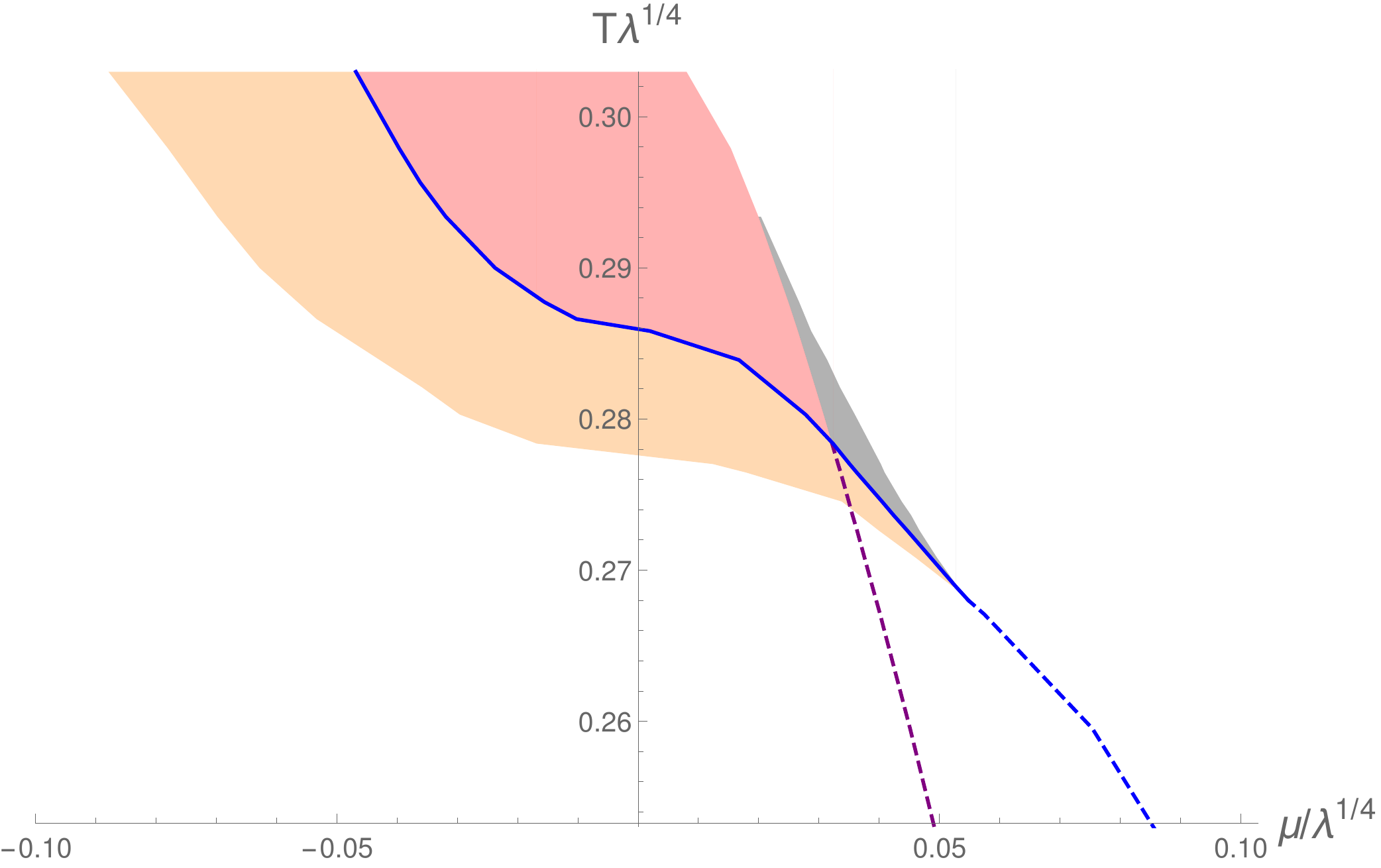}  }  \vspace{0.3cm}\\ 
\mbox{\includegraphics[scale=0.34]{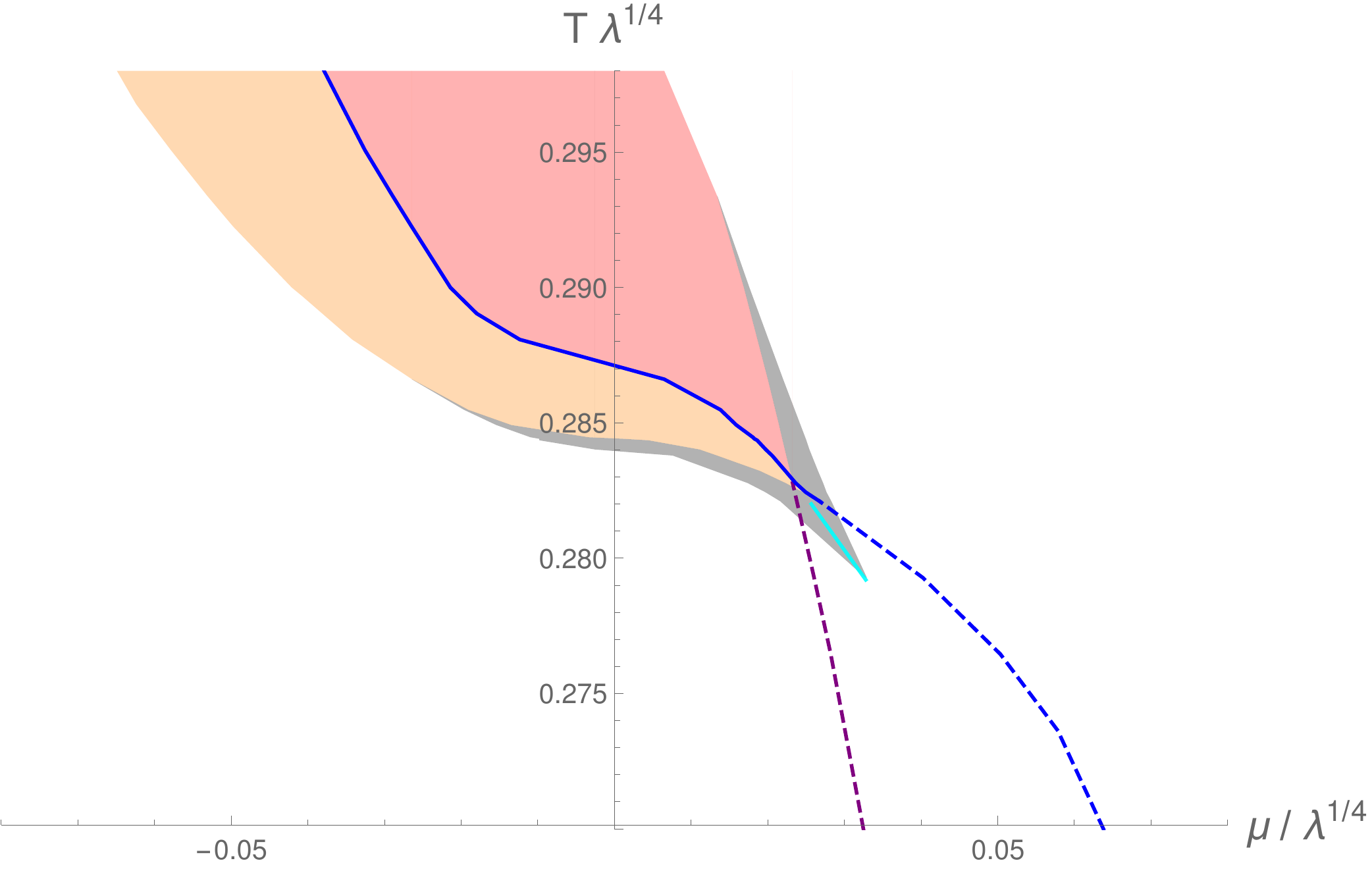}  \hspace{1.7cm} \includegraphics[scale=0.34]{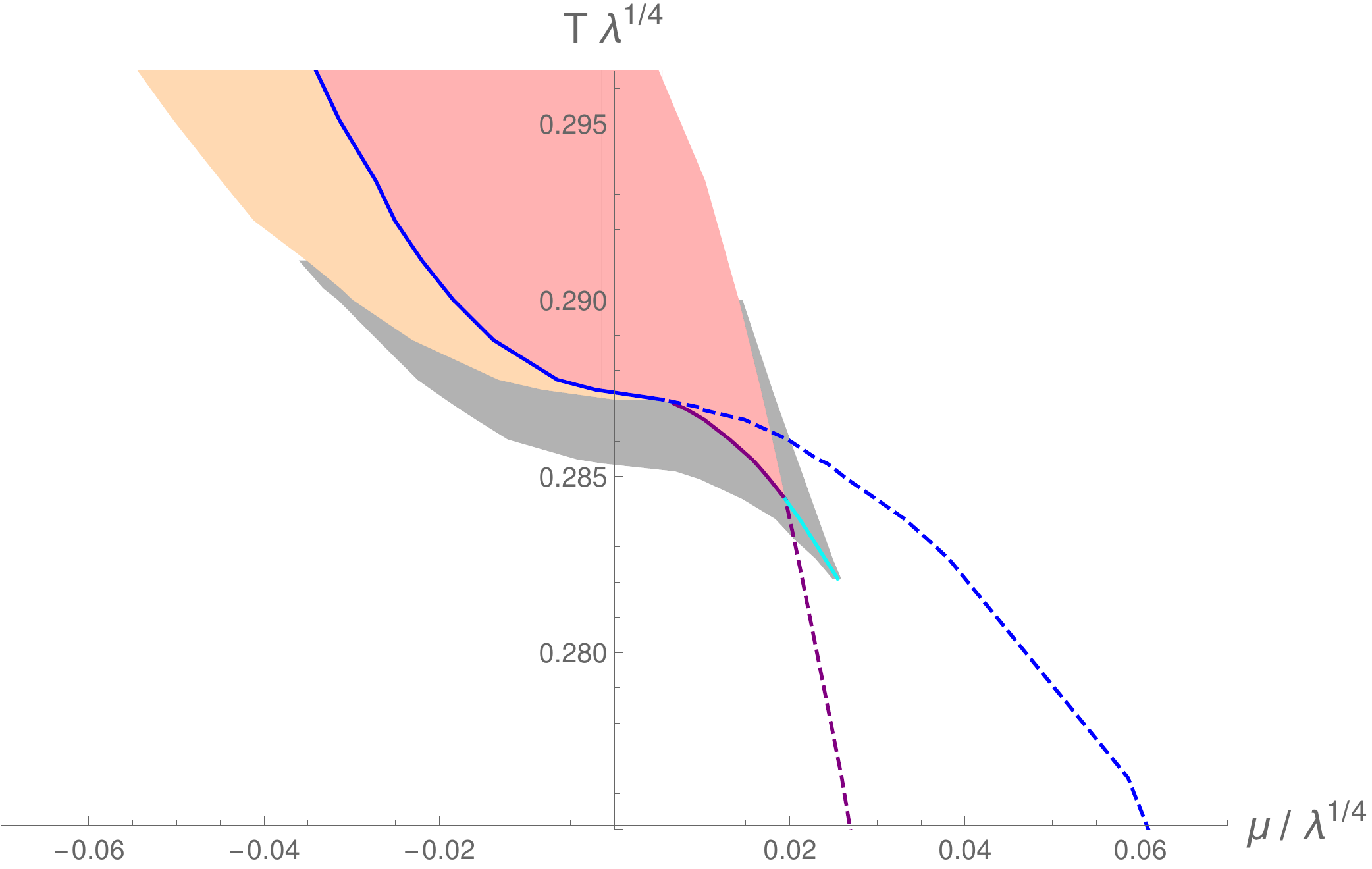} } \vspace{0.3cm} \\ 
\mbox{\includegraphics[scale=0.34]{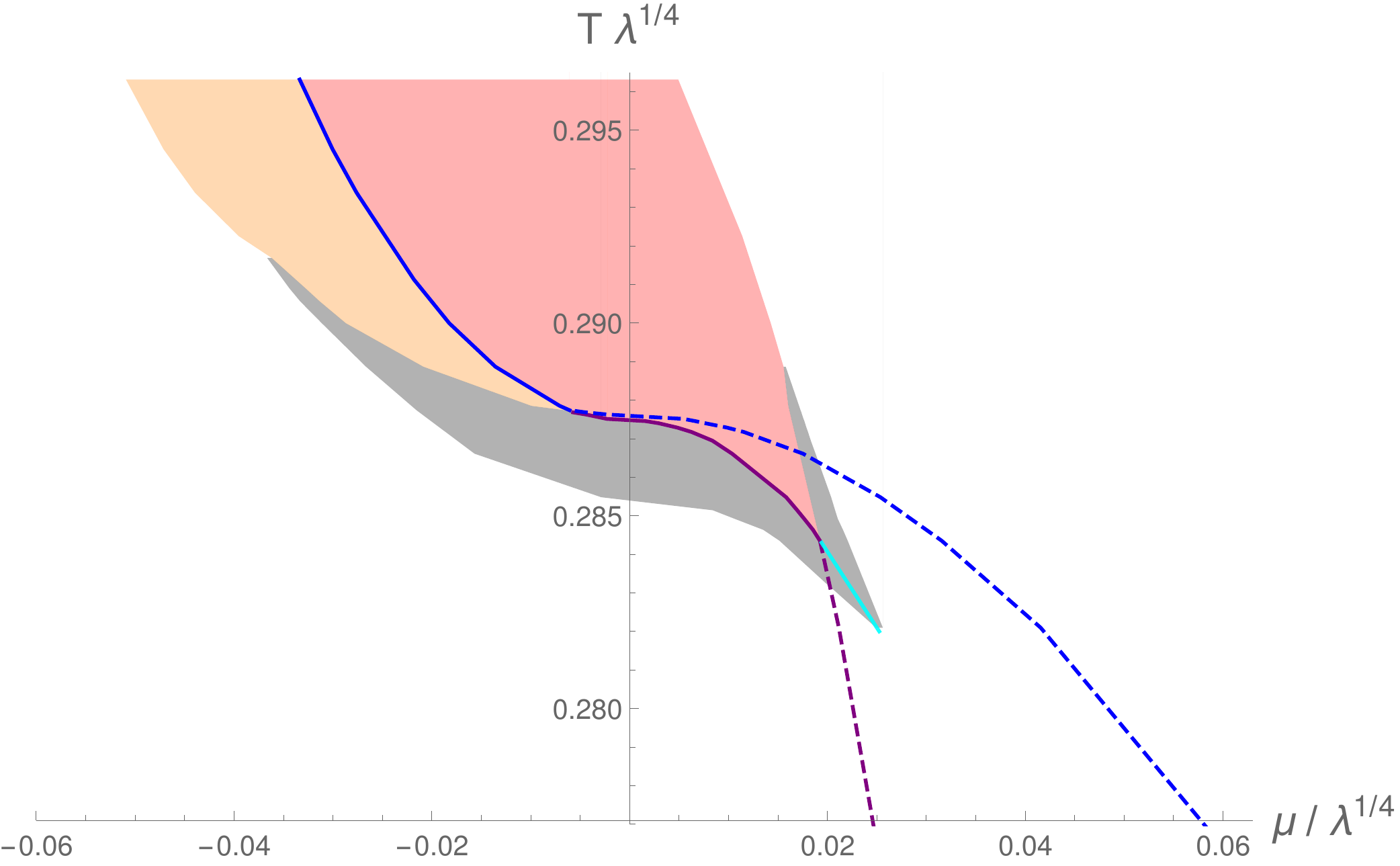}  \hspace{1.5cm} \includegraphics[scale=0.34]{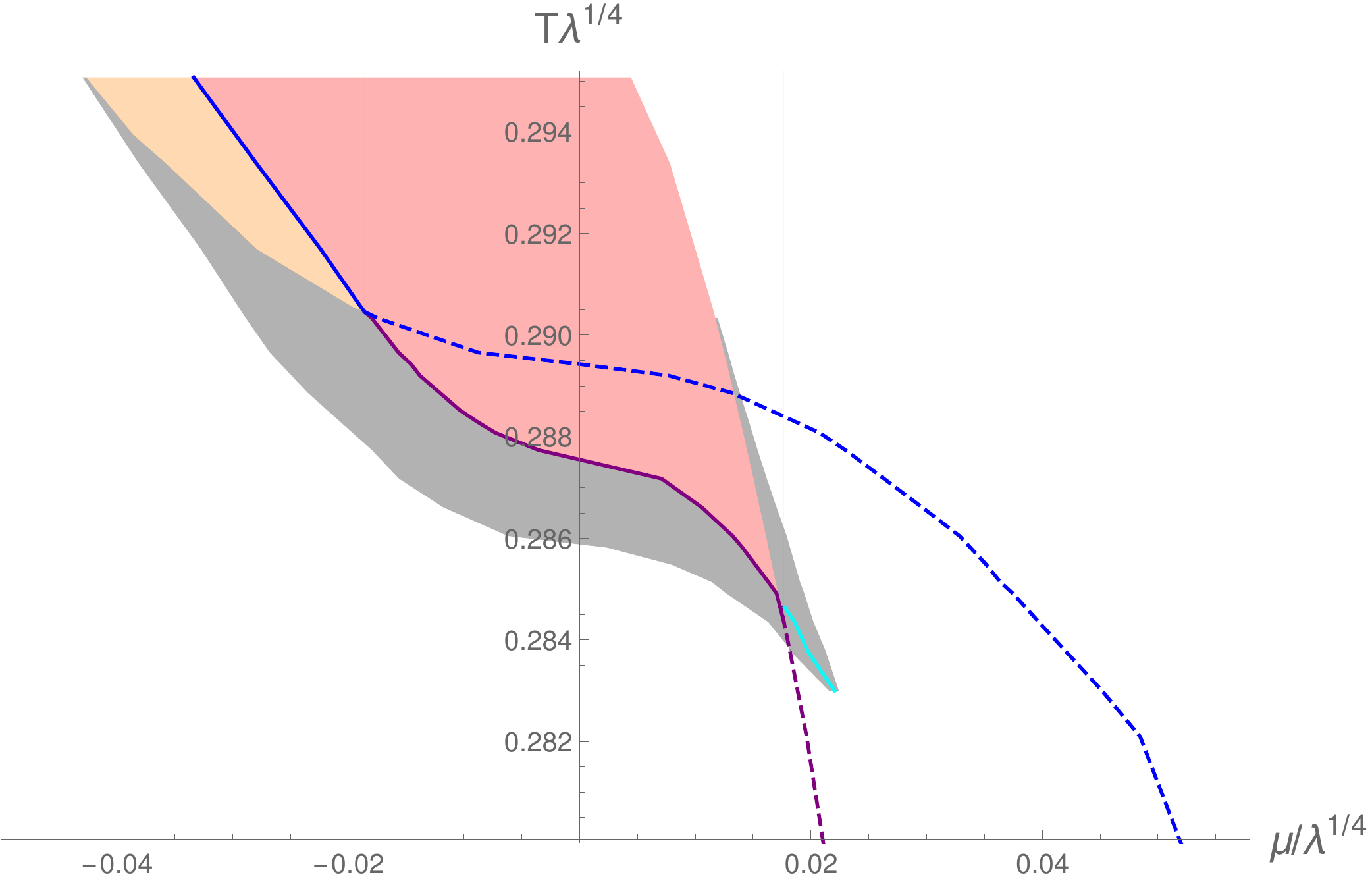} }  \vspace{0.3cm}\\ 
\includegraphics[scale=0.34]{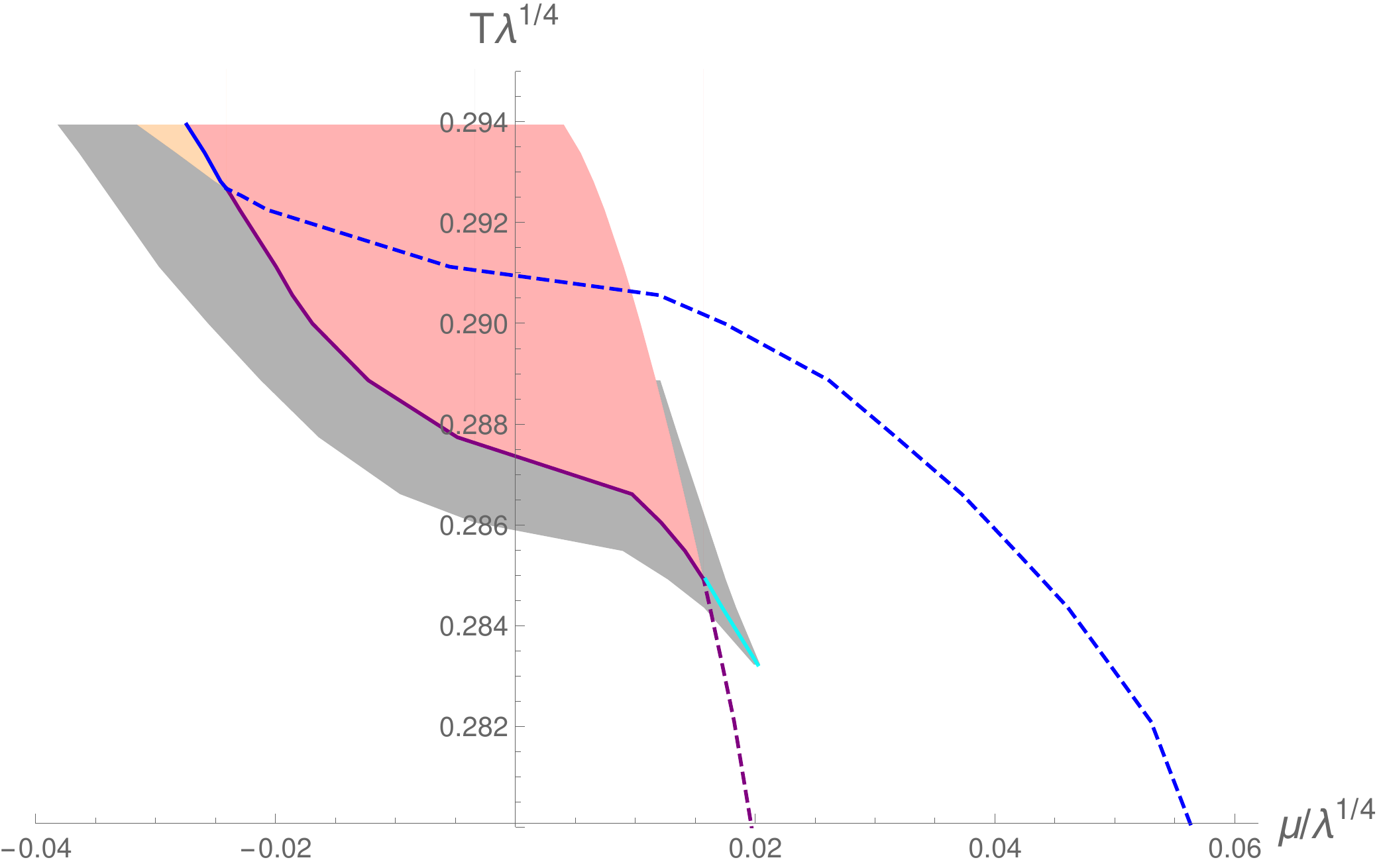}    

\mbox{ Figure IV: $T-\mu$ phase diagrams for the  deformed theories with $\alpha$ = 1,~1.08,~1.12,~1.137,~1.139, ~1.15, ~1.16.
}
\label{slices}
\end{center}   

\newpage

$\left. \right.$ \newpage

\begin{center}
\includegraphics[scale=0.5]{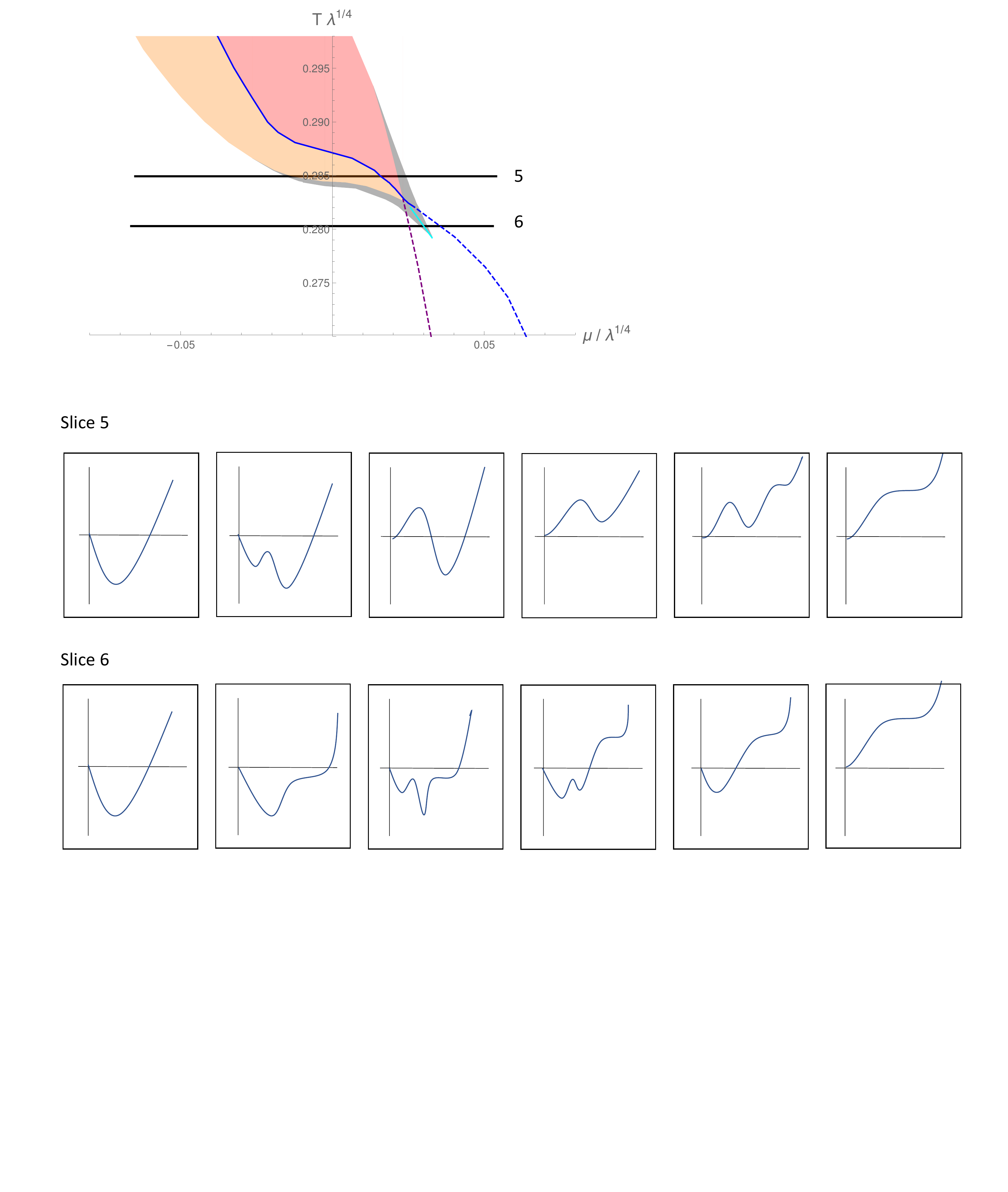}  \vspace{-5cm}

\mbox{ \it Figure V:  The effective potential evolution (vs $c$) from left to right as one moves along two T slices at $\alpha$ = 1.12.}
\end{center}

 density onset transition. By $\alpha = 1.12$ the critical points on both these lines have moved together. At larger $\alpha$ the density line shows a period of first order behaviour  after separating from the chiral transition line at the critical point.  

Previous analysis \cite{Evans:2011eu} had seen that for big enough $\alpha$ one could make the $\mu = 0$ chiral transition second order but 
now we see that in fact we have pushed the critical point into the imaginary $\mu_I$ plane. It's fascinating to think that some deformation might do this in QCD itself!

The region of metastable vacua broadly speaking moves with the points but there are some new features which we highlight by sketching the effective potential 
for a couple of slices in the $\alpha = 1.12$ case - see Fig V. The first slice

\newpage

$\left. \right.$ \vspace{16.6cm}

just highlights the meaning of the grey zones around the edge of the full metastable region. Moving along slice 5 from the left to right the chirally symmetric embedding must become metastable from initially being a potential maximum. Previously it did this by casting out another maximum. Here though  
a maximum and a minimum pair create away from the chirally symmetric solution (both black hole solutions) and the minimum then merges with the chirally symmetric vacuum to convert it to a metastable minimum. Similarly after the first order transition the chirally broken remnant vacuum does not simply merge with the potential maximum but casts off another minimum (becoming a point of inflection) that annihilates the maximum. Thus in the grey zones there are metastable vacua with a massive deconfined quark phase.  Since these are minima created between the previous maximum and minimum of the potential (the chirally broken and chiral restored vacua) the metastable vacuum is typically not very deep which suggest it would be harder to find either on the lattice or in heavy ion collision events.

Slice 6 shows a remnant of these ``pair creation'' events forming a ``fishtail crossover'' between the two second order transitions. Here after the chiral symmetry breaking vacuum has second order transitioned to a massive deconfined quark phase a maximum and minium pair create. There is then a first order transition (the cyan line) between two massive deconfined quark phases. Finally the metastable vacuum annihilates with the maximum to allow a second order transition from the massive deconfined quark phase to the chirally symmetric phase. This 
explains the tail transition between the legs in these plots (which we view as a minor part of the story). 

Let us again comment on the broad picture and lessons. The lines that mark the borders of where the chirally symmetric or chirally broken phases become metastable continue to play a crucial role. In all these phase diagrams these borders  plus the first order phase line itself join at the critical points -  in each case the two lines point at the critical point. Further these borders are distinct around the first order transition and then cross as they become the second order lines. Note even in the case where the two borders both cross at the critical point ($\alpha = 1.12$) they do not merge but separate again. This 
is the origin of the deconfined massive quark phase.

None of the cases we have seen have second order transitions at $\mu=0$ and first order at $T=0$ as expected in QCD. However, the cases where the critical point
 lies in the imaginary $\mu_I$ plane can allow us to speculate. Imagine now a putative theorist who (somehow) can only compute at real $\mu_R$ but not imaginary $\mu_I$. In these cases the theorist would just see a second order chiral transition. In this model, however,  if they could identify both the denisity onset transition and the chiral transition then those lines can be extrapolated to the critical point in the $\mu_I$ plane where he is ignorant.

\section{General Lessons \& Questions for the QCD Phase Strcuture}

So far it has been interesting as a purely theoretical problem to investigate the structure in the phase diagram
 (extended to imaginary $\mu$) of an exactly solvable gauge theory and amusing to look for signals of transitions and critical points in one part of the plane if one only had access to a sub-region. Have we learnt
any lessons that 
could be applied more widely to a generic set of chiral symmetry breaking models (perhaps at large $N_c$) or even to QCD? We have concluded that looking for the regions of the phase diagram with metastable 
vacua can be used to identify the 
 positions of the chiral critical point. Using this insight, in Figure VI we propose a number of qualitative pictures for theories where the transition is first order at at high $\mu_R$ and low T, but second order transitions at higher T and lower $\mu$ or at $\mu_I$.

Let us begin by simply talking about theories without confinement that seem a natural extrapolation of the ones we have studied and ask in these worlds what lattice or 
 heavy ion data could reveal. 
The first sketch in Figure VI shows the most pessimistic possible conclusion. Here 
we assume that the second order transitions (chiral cross over at finite mass) for density switching  on and chiral restoration are degenerate and that the first order transition exists only at large real $\mu$. The metastable region could be quite tightly positioned around the first order transition. Here there is 
little hope of using the lattice to idenitfy anything beyond the position of the second order line which provides no information on the position of the critical  point. Here one might hope to use heavy ion collisions at low $T$ to identify $T, \mu$ points within the region with metastable 
vacua (again assuming that events that get stuck in the metastable vacua can be distinguished after hadronization). 

The second sketch is a more hopeful speculation where the region with metastable vacua might be wider about 
the first order line - here one could hope to find regions of metastability at low T and low $\mu$ or even imaginary $\mu$ on the lattice. The second order phase transition line and the edge of the region of metastability converge at the critical point and could be used to point to it.  Here we have also allowed a boundary region around the metastability region with a metastable massive deconfined quark phase. Tracking from left to right across the metastability region would have an effective potential that changes as in slice 5 of Figure V. Here the key question is whether the chirally symmetric vacuum converts from a maximum 
to a minimum directly by spitting out a maximum or whether a maximum and minimum are pair created elsewhere in the potential with that minimum then joining to the chirally symmetric vauum to make it a minimum. A priori both seem possible.

The third and fourth sketches show the structure one would expect if the chiral and density transitions separate. The critical line and one boundary of the metastable region meet at each critical point so could be used to predict its position. 

Finally let us tentatively speculate for QCD. The first additional issue we must consider is confinement that is not included in the model we have used. We already know that at T=0 the first transition with $\mu$ is the first order switch on of baryon number. This transition might be distinct from the deconfined quark pictures we have drawn so far in which case the first 4 sketches could all

\newpage

\begin{center}
\mbox{\includegraphics[scale=0.25]{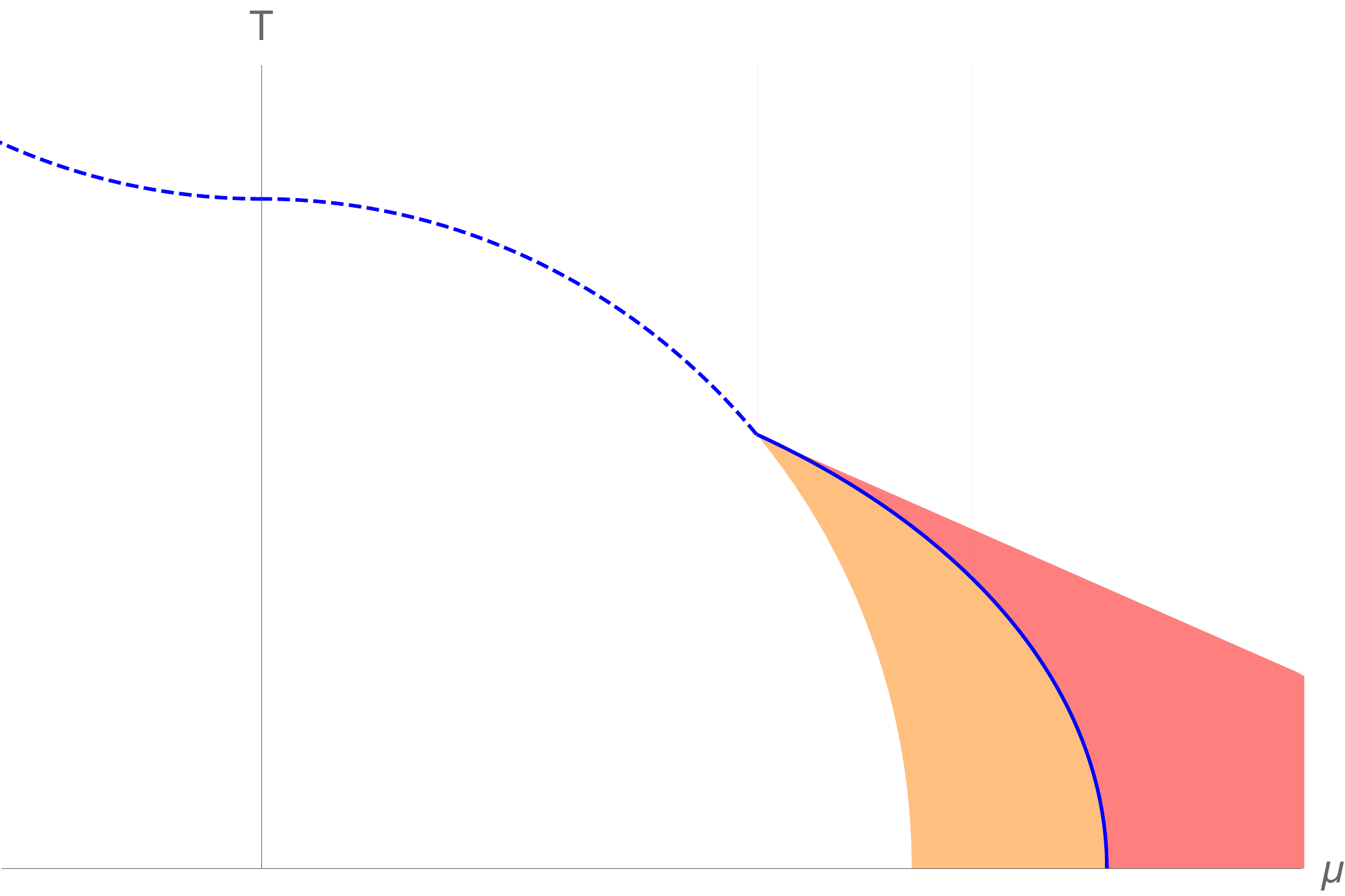} \hspace{0.3cm} \includegraphics[scale=0.25]{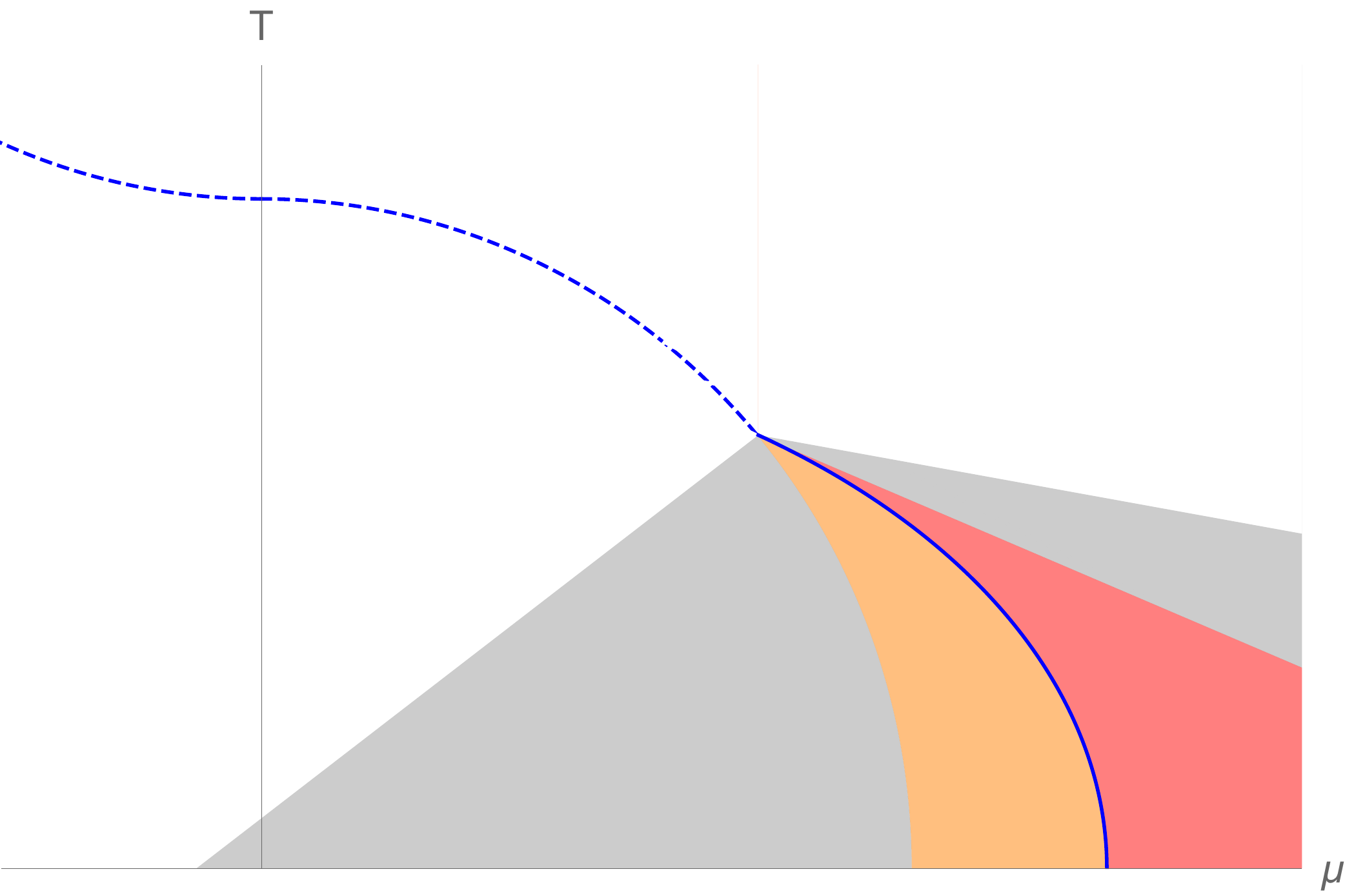}  
\hspace{0.3cm} \includegraphics[scale=0.25]{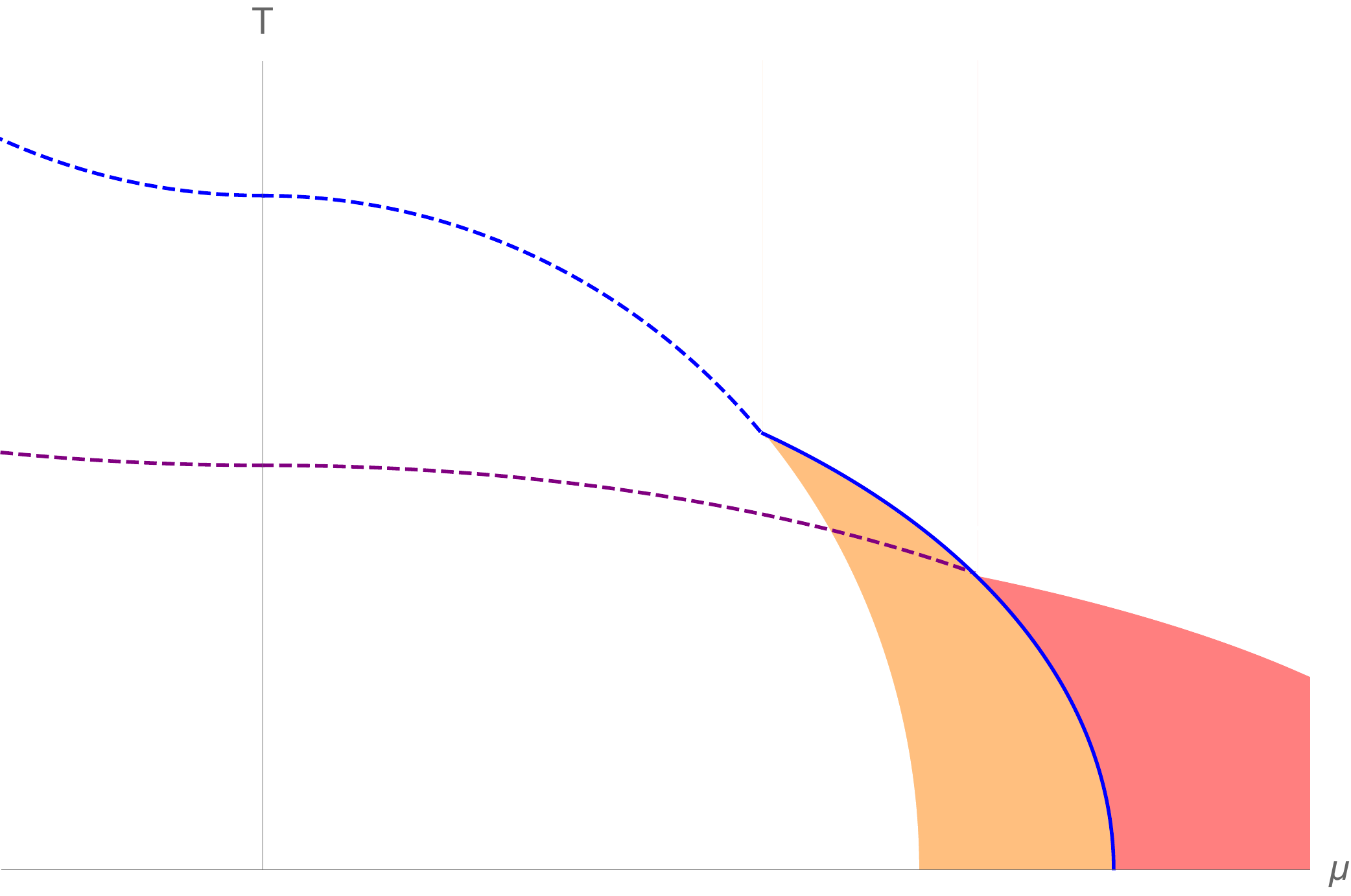}  }  \vspace{0.5cm}

\mbox{\includegraphics[scale=0.25]{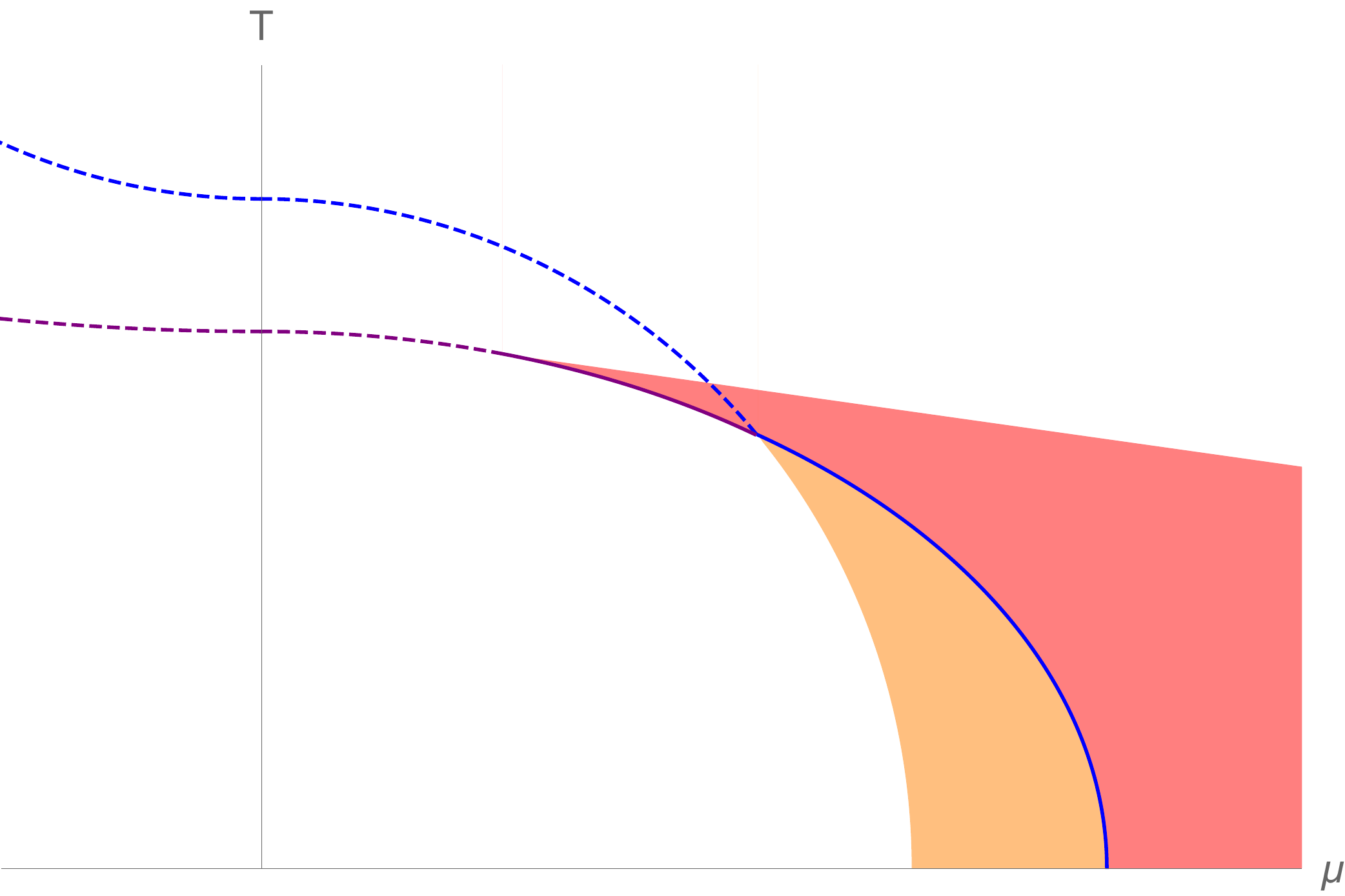} \hspace{3cm} 
\hspace{0.3cm} \includegraphics[scale=0.25]{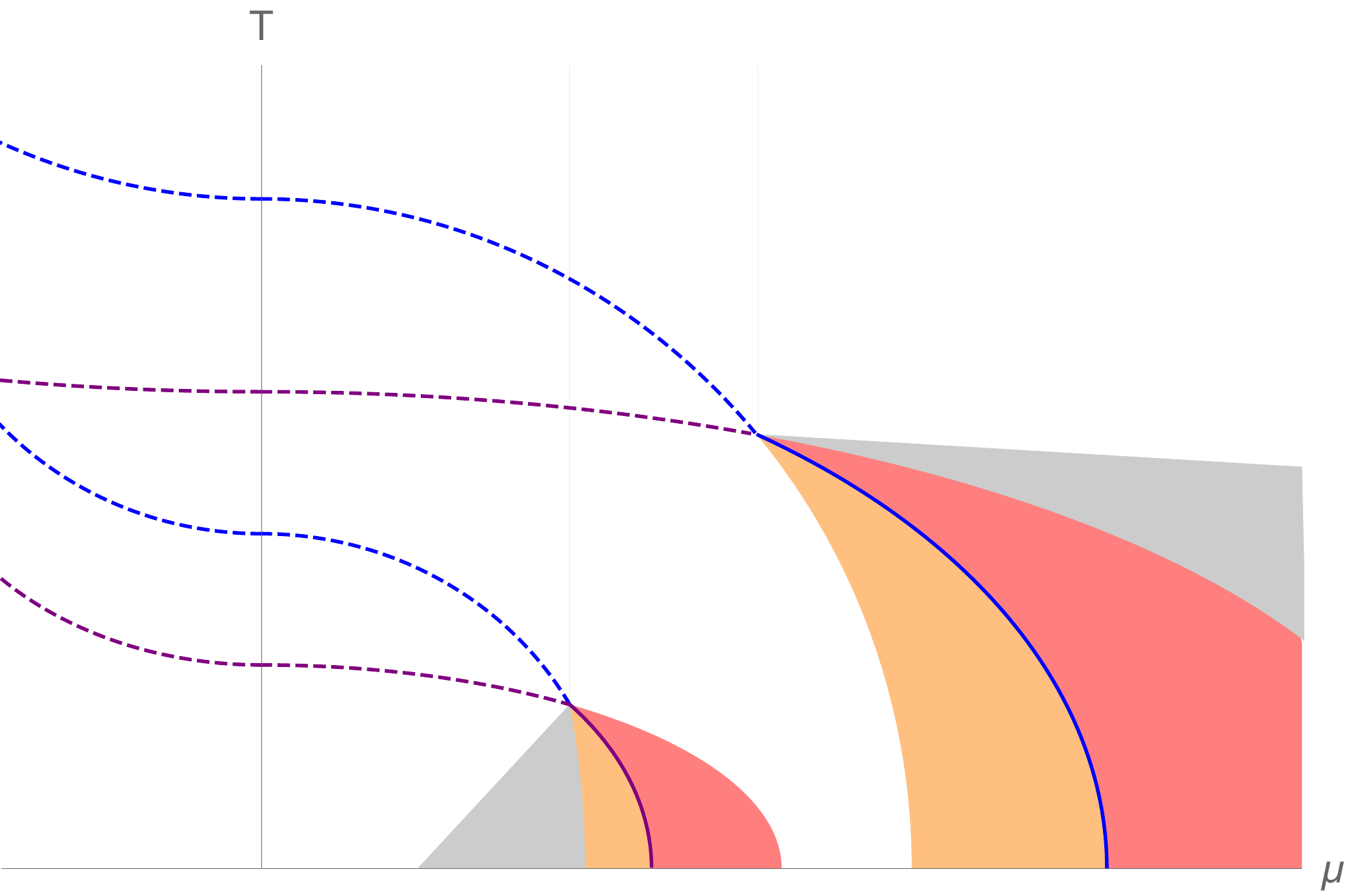}  }

    \end{center}  

\mbox{ \it Figure VI:  Here we present some speculative sketches for the phase diagrams of theories with first order transitions}
\mbox{\it  at high $\mu_R$ and low T, but second order transitions at higher T and lower $\mu$ or at $\mu_I$. Here dashed lines are second }
\mbox{\it order  transitions; solid lines are first order transitions; in the red region the chiral symmetry breaking vacua is }
\mbox{\it metastable; in the orange region the chirally symmetric vacuum is metastable; in grey regions dense massive quark}
\noindent \mbox{\it  phases are metastable. The final sketch shows the case that might apply to QCD where the
 density transition is}
 \mbox{\it  linked to the known nuclear  density onset and the chiral transition is separate (here the outer red and organge}
 \mbox{\it  regions are metastable dense massive quark vacua).} 

lie  to the right of the baryon onset transition. However, it also seems natural to associate the baryon density transition with the bottom of the density phase transition we have seen. The chiral transition is then separate but
potentially also first order. We sketch such a set up in the final picture of Fig VI. Here we appear to have drawn a deconfined, dense  but chirally broken phase between two phase boundaries all along the transition. On the other hand we know that on the left hand boundary at low T these quarks should be confined and we should treat this as the baryonic phase. It is possible that within this region there is a transition where confinement switches off and a deconfined massive quark phase is realized (as  speculated in \cite{Fadafa:2019euu}), but equally confinement may cover the whole phase region.  

Note in this final picture the left hand red region is where the chirally broken vacua is metastable and the right most orange region where the chirally symmetric vacuum is metastable. The orange and red regions on the outer edges, in the language of our brane model, would be metastable dense and chirally broken vacua. If we believe our structures then these boundaries would continue beyond the critical region as further second order boundaries between a variety of dense yet chirally broken vacua. In QCD the full region between density switching on and chiral symmetry being restored is the cross over region. Our model suggests there might be further 
 
\newpage

$\left. \right.$ \vspace{11.6cm}

 second order transitions within that cross over region! 
  In reality in QCD these are likely to be smoothed to cross overs and be
 very hard to spot if they exist at all. On the other hand at finite $\mu$ this cross over region might widen and allow more structure to be spotted (Fig 4 in \cite{Bellwied:2015rza} suggest the cross over region may widen at larger $\mu_R$). 
 
 Finally we can again speculate that metastable vacua of some sort might exist over a wide region of the low T phase diagram that could be hunted for on the lattice at low $\mu_R$ or even at $\mu_I$ or that might display as
 new types of event in heavy ion collisions. Of course both are difficult and expensive technologies to use for such speculative searches!

\section{Summary}

The AdS/CFT Correspondence allows the exact computation of the phase diagram of an ${\cal N}=2$ gauge theory with a small number of quark hypermultiplets in the fundamental representation in the presence of a magnetic field which triggers chiral symmetry breaking. The phase diagram was computed previously in \cite{Evans:2010iy} and is
 shown on the right in Fig I. There are regions where the chiral restoration transition are first order, second order and there is a linking critical point. Here we first extended the phase diagram to imaginary chemical potential, $\mu_I$, shown on the left hand side of Fig I. Only a small region of the $\mu_I$ plane is stable and there the first order transition extends from the real $\mu_R$ plane. 

As shown in Fig I it is hard to imagine deducing anything about the phase structure at $\mu_R$ from the $\mu_I$ plane. However, we have added to the figure regions in which there are metastable vacua - see Fig II. Amusingly the edges of these regions and the phase line itself form arrows pointing to the critical points in the $\mu_R$ plane.  One could hope to identify the edges of these arrows in the $\mu_I$ plane or at low $\mu_R$ and extrapolate to find the critical points approximate position.

If one approaches the critical points from large $\mu_R$ in this theory the second order transition line in fact splits into a line at which density switches on and another at which the chiral condensate swiches off. Here we have understood that these second order transition lines are the natural extensions of the boundaries of the region where there are  metastable vacua around the first order transition line. It seems very natural that those lines should remain distinct in the second order region. The two second order transition lines again converge at a critcal point.

We have also explored variants of these phenomena in a bottom up deformed version of the theory which allows the critical point to be pushed into the complex chemical potential segment of the plane. See Fig IV. 

These observations have led us to speculate about QCD. Could the cross over region from the onset of density to the restoration of chiral symmetry actually contain several distinct transitions or crossovers corresponding to the continuation of the edges of the metastable regions needed in QCD? At imaginary chemical potential this separation might be wider than on the T axis. If these seperate transitions could be identified then they could be used to predict the position of the critical point. Secondly it would be worth searching the rest of the accessible $\mu-T$   
plane for metastable vacua since the boundaries of these regions also contain information about the position of the critical point. Of course there is no guarantee that these features are present or identifiable on the lattice or using heavy ion data but they are intruiging possibilities. \bigskip

\noindent{\bf \large Acknowledgements: } We thank Prem Kumar, Andreas Schmitt, Keun-Young Kim and Gert Aarts for fruitful discussions. NE’s work was supported by the STFC consolidated grant ST/P000711/1 and MR by an STFC studentship. \vspace{1cm}

\begin{center}
	\includegraphics[scale=0.38]{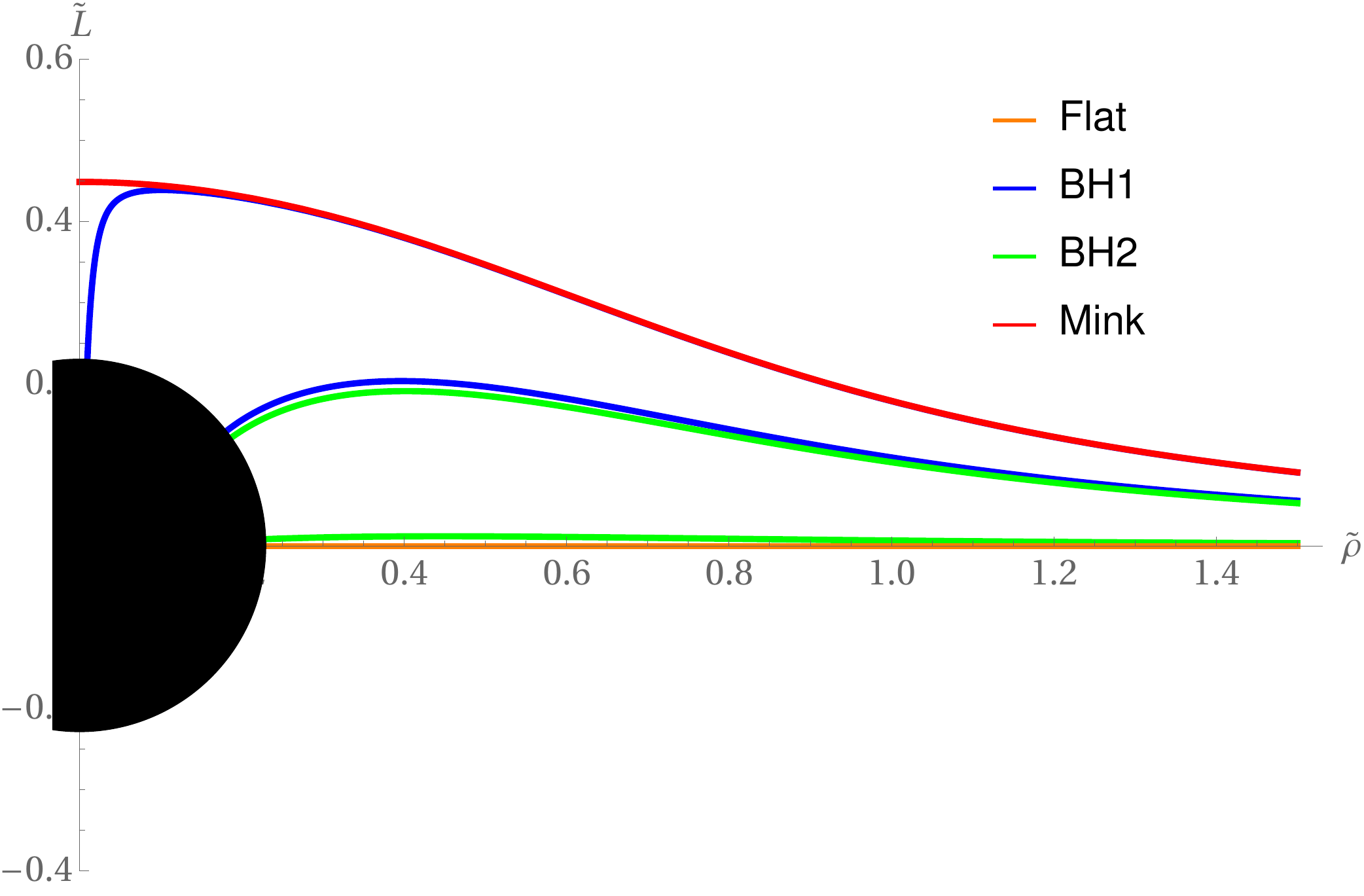}    
	\includegraphics[scale=0.38]{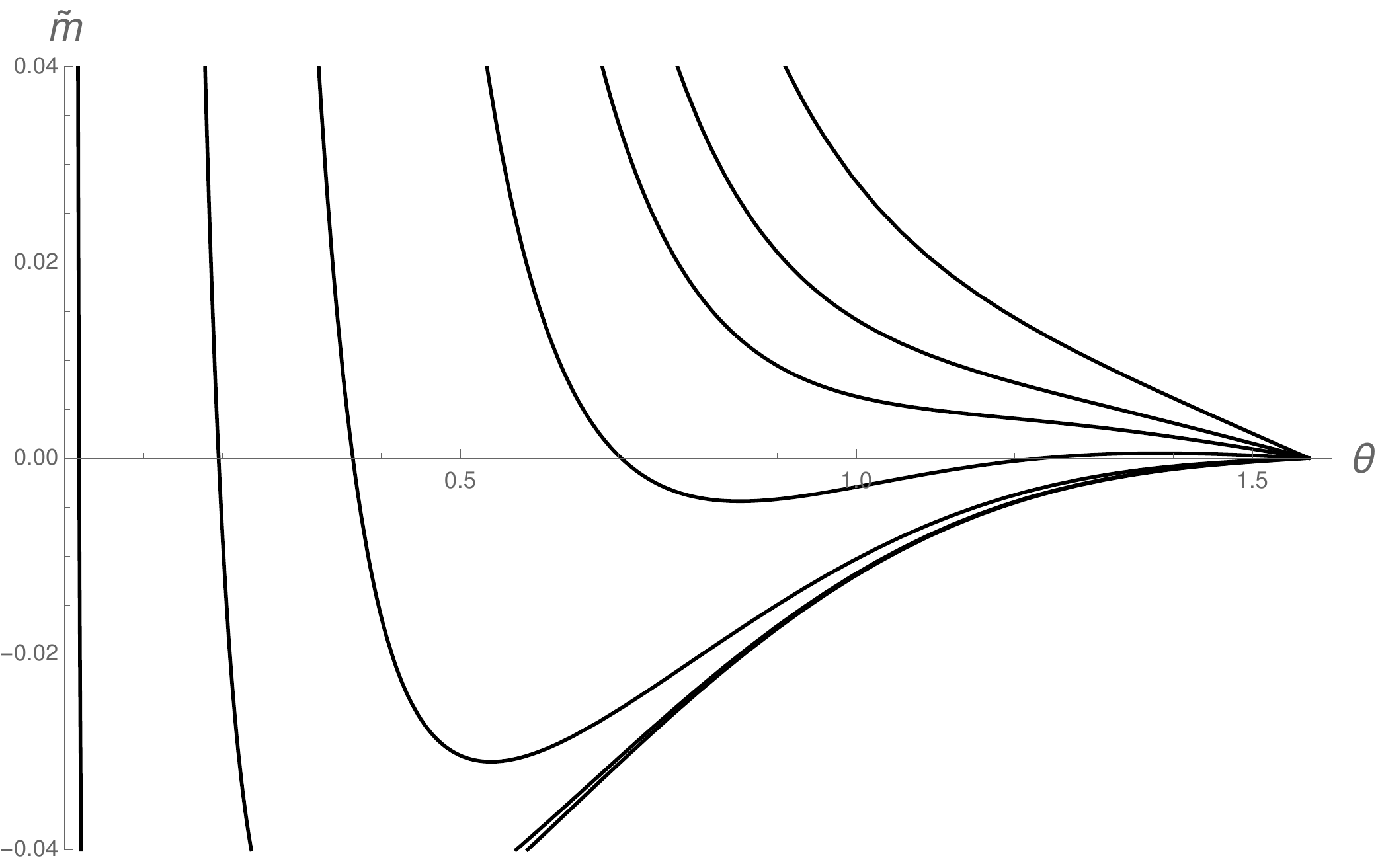}
	\end{center}
	{\it FIGURE VII: Detailed plots for the $\alpha=1$ theory with T=0.26. \\ (i)The top plot shows sample embedding functions (red Minkowski, Orange flat, BH1 in green and BH2 in blue.  \\ (ii) the second plot shows the UV mass of embeddings emerging at angle $\theta$ from the black hole as a function of $d$ (for values $d=$0.001, 0.01, 0.03,  0.07, 0.1, 0.12, 0.15 from left to right\\
	}

	\begin{center}
{\bf APPENDIX A: COMPUTATIONAL DETAILS FOR A SINGLE T SLICE}
\end{center}

Here we present detailed plots of the $T=0.26$ slice across the phase diagram in the $\alpha = 1$ theory - slice 4 in Fig III.  This provides more detail on how we construct the phase diagrams we have presented.

We plot example D7 embeddings in the top figure in Fig VII. In red is the Minkowski embedding with $\tilde{d}=0$; orange is the flat embedding $\tilde{L}=0$ with $\tilde{d} \neq0$. In between we plot some black hole embedings. These are found from the second plot down in Fig VII - here at a given $\tilde{d}$ we shoot off the horizon at different angles
	$\theta$ and plot the resulting UV mass value $\tilde{m}$ - we seek solutions with $\tilde{m}=0$. As can be seen between $\tilde{d}=0.03$ and $\tilde{d}=0.07$ the number of solutions change as the minimum
	
	\newpage

\begin{center}
	
	\includegraphics[scale=0.38]{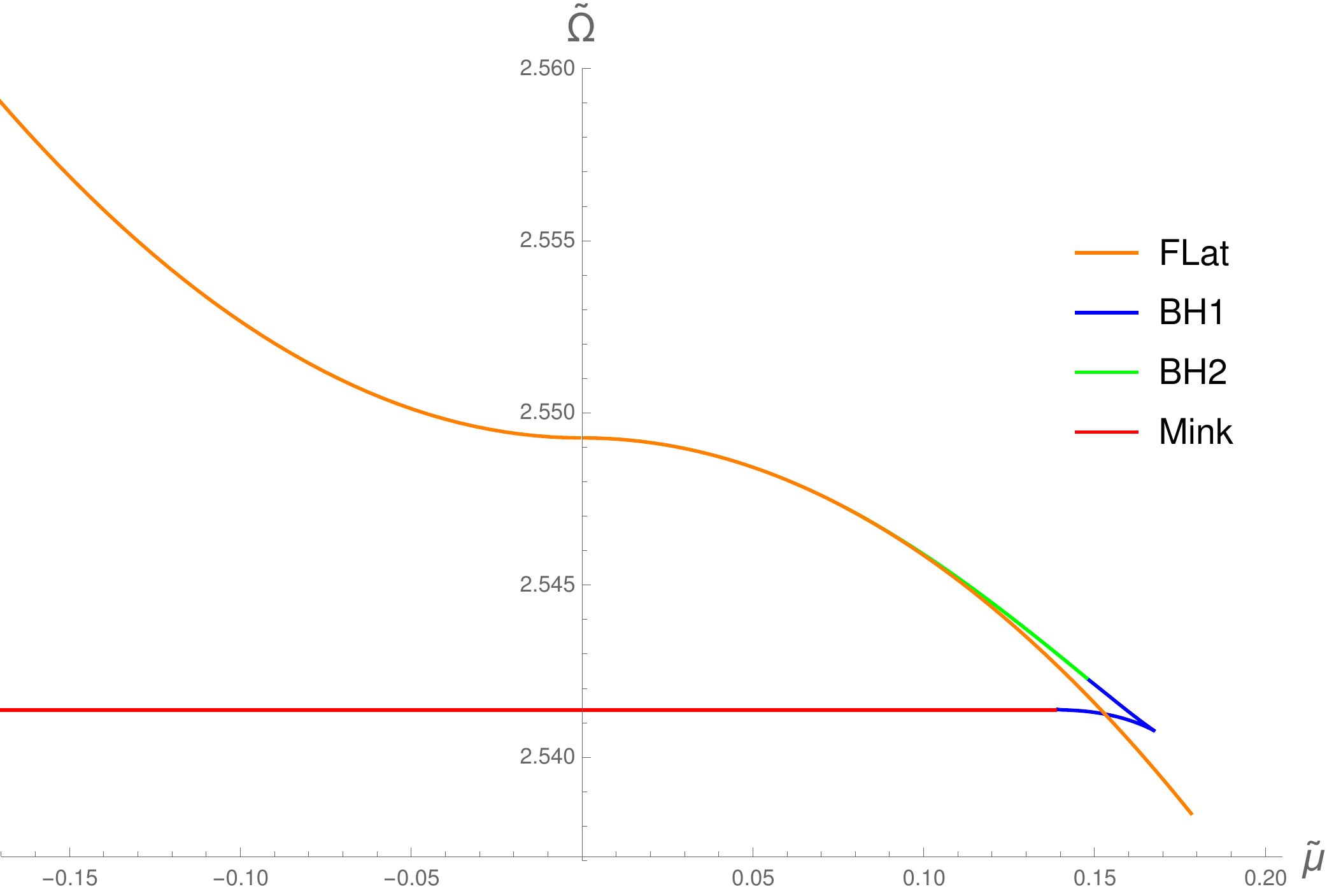}
	\includegraphics[scale=0.38]{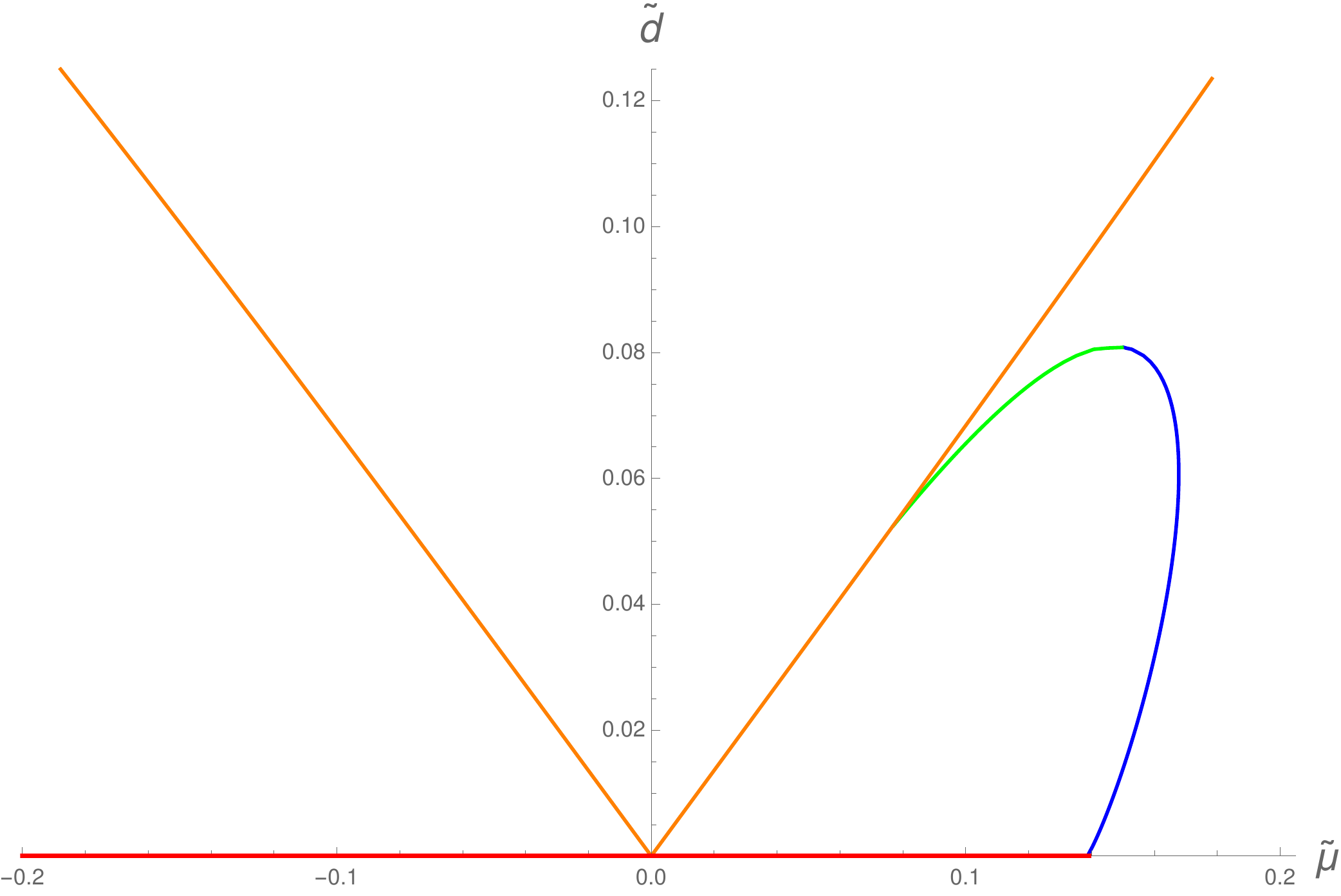}
	\end{center}
	{\it FIGURE VIII: Detailed plots for the $\alpha=1$ theory with T=0.26. 
	(i) the third plot shows the free energy of the solutions against $\mu$\\
	(ii) the fourth plot shows $d$ vs $\mu$ for the solutions.}

  of the curve passes through the $\tilde{m}=0$ axis. This corresponds ot the pair creation of two black hole solutions which we will call BH1 and BH2 - the two solutions emerge as the closest green and blue black hole solutions in the top plot. By following the evolution of the two solutions in the second plot it can be seen that one moves to merge with the  Minkowski embedding and the other with the flat embedding - the outer two blue and green embeddings in the top plot.

Now we can, embedding by embedding, compute $\tilde{\mu}$ from (\ref{mud}) and the free energy $\tilde{\Omega}=-{\cal \tilde{L}}$ in (\ref{muL}) . We plot $\tilde{\Omega}$ against $\tilde{\mu}$ for our solutions in the third plot  and $\tilde{d}$ versus $\tilde{\mu}$ in the bottom plot in Fig VII.

The free energy plot allows us to clearly see the phase structure along the slice, from left to right. At imaginary $\tilde{\mu}_I$ the Minkowski embedding is the lowest energy state and the flat embedding is the maxiumum of the potential. The first transition is where a BH2 solution emerges from the flat embedding - the flat embedding has become a local minimum of the potential.  Next a BH1 solution emerges from thre Minkowski embedding and has lower energy - there is second order transition to the BH1 state as density switches on (we do not plot the continuation of the Minkowski embedding red line in the plot   
further to the right although it does continue to exist). There is then a first order transition from the BH1 embedding to the flat embedding. The BH1 state is a metastable state briefly although with energy quite near the flat embedding of the true vacuum. Finally the BH1  solution ceases to exist merging with a BH1 solution that is the continuation of the BH2 state that is the local potential maximum. Note that the annihilation  of the BH1 and BH2 solution as identified in the density plots is an innocuous transition when plotted with $\tilde{\mu}$.

\end{document}